# GENDER DISPARITIES IN INTERNATIONAL RESEARCH COLLABORATION: A STUDY OF 25,000 UNIVERSITY PROFESSORS

Marek Kwiek*

*Center for Public Policy Studies, University of Poznan, Poland*

Wojciech Roszka

*Institute of Informatics and Quantitative Economics, Poznan University of Economics and Business*
*Poznan, Poland*

**Abstract.** In this large-scale research based on bibliometric, biographical and administrative data, we examine how gender disparities in international research collaboration differ by collaboration intensity, academic position, age, and academic discipline. The following are the major findings: (1) While female scientists exhibit a higher rate of general, national, and institutional collaboration, male scientists exhibit a higher rate of international collaboration. (2) An aggregated picture of gender disparities hides a more nuanced cross-disciplinary picture of them. (3) An analysis of international research collaboration at three separate intensity levels (low, medium, and high) reveals that male scientists dominate in international collaboration at each level. However, at each level, there are specific disciplines in which females collaborate internationally more than males. Furthermore (4), gender disparities in international research collaboration are clearly linked with age: they are the lowest and statistically insignificant for young scientists and the highest and statistically significant for the oldest scientists. Finally, we estimate the odds ratios of being involved in international research collaboration using an analytical linear logistic model. The examined sample includes 25,463 internationally visible Polish university professors from 85 universities, grouped into 24 disciplines, and 158,743 Scopus-indexed articles.

## 1. Introduction

In the study of science, gender is “a strategic research site”: gender hierarchies are still “pervasive” in science (Fox, 2020, p. 1001). Gender tends to influence research performance, academic rank, and citation impact in academic science, both for whole national populations of scientists and their apex, national research top performers (Larivière *et al*., 2011; Abramo *et al*., 2019). Moreover, and most importantly for this paper, "collaboration patterns vary by gender" (Bozeman *et al.*, 2012, p. 8). The relative rate of collaborative research has increased over time, and it currently dominates solo authorships in all fields, except for the humanities (Wuchty *et al.*, 2007). International research collaboration is a hot topic in policy studies on the increasing globalization and networking in science (Wagner, 2018). Furthermore, females in science (and tackling discrimination against them through various equality strategies) is another popular policy topic (Zippel, 2017)—with significant policy implications—which makes gender disparities in international research collaboration a highly interesting theme, particularly if large-scale data encompassing entire national systems are utilized, as in the case of this study. However, as Abramo *et al.* (2013) note, the debate on gender aspects in research systems has focused primarily on the overrepresentation of male academics, the productivity gap, and gender discrimination, and only rarely on collaboration patterns. In this research, we examine the hypothesis that gender disparities in collaboration patterns in science differ in terms of collaboration intensity, academic position, age, and discipline, thereby reflecting wider gender disparities in all collaboration types (general, international, institutional, and national). In addition, we estimate the odds of being involved in international research collaboration using an analytical linear logistic model.

*Corresponding author contact email: kwiekm@amu.edu.pl.

An integrated data set of all Polish scientists with their administrative, biographical, publication, and citation data is used in this study (the Polish Science Observatory data set maintained by the authors includes 99,535 scientists and 377,886 Scopus-indexed publications published in the decade 2009–2018). The sample examined in this paper comprises 25,463 internationally visible Polish university professors from all 85 universities with internationally visible publications within the decade, grouped into 24 disciplines, and 158,743 Scopus-indexed journal articles. The individual scientist, rather than the individual article, is the unit of analysis. The key methodological step, unique in studies of collaboration patterns, is the determination of what we term an "individual publication portfolio" (for the decade of 2009–2018) for every internationally visible Polish scientist.

The paper is structured in the following manner. The next section provides a literature review, followed by data and methods. The "Results" section presents discussions on gender disparities in international collaboration from the perspectives of collaboration intensity, academic disciplines, age, and academic positions, as well as on the results of the linear logistic model. The last section ends the paper with a summary of the findings, discussion, and conclusions.

## 2. International Research Collaboration and Gender: Literature Review

### 2.1 *Changing Global Contexts: Time Matters*

In the specific European context (de Wit and Hunter, 2017, p. 25; Hoenig, 2017; Olechnicka *et al.*, 2019, pp. 140–145), international research collaboration and international mobility define academic career prospects and determine individual and institutional access to national and European research funding. The link is straightforward: within the European Research Area, international research collaboration and international mobility have become a metric of excellence and quality (Fox *et al.*, 2017, p. 1293); they have been prioritized and funded as critically important in the past two decades, being closely linked with European integration and European Union research funding (which generally requires research partners from at least three countries; see Mattsson *et al.*, 2010, p. 556). While scientists vary in their individual predilection to collaborate and coauthor internationally (Glänzel, 2001, p. 69), "the more elite the scientist, the more likely it is that he or she will be an active member of the global invisible college" (Wagner, 2008, p. 15; see Kwiek, 2016 on the European research elite). International research collaboration and international mobility across Europe increases scientists' chances of securing an academic position (Herschberg *et al.*, 2018, pp. 822–823), moving faster up the career ladder (Fochler *et al.*, 2016, pp. 196–198), and securing external funding for their research, including highly prestigious funding from the European Research Council, promoting scientific “excellence” at a supranational level (Hoenig, 2017, pp. 3–5). “Internationalization” and scientific “excellence” come together in socially stratified European science (Kwiek, 2019). International research collaboration is reported as strongly influenced by academic discipline, institutional type, and national reward structure (Cummings and Finkelstein, 2012, p. 86; Finkelstein and Sethi, 2014, p. 235; Kyvik and Aksnes, 2015, p. 1442), and it is a defining feature of a new global geography of science (Olechnicka *et al*., 2019). However, the benefits of collaboration, rather than being constant, vary internationally, and collaboration strategies that are effective in one nation are not easily transferable to another (Thelwall and Maflahi, 2019, p. 13). While gender disparities in research collaboration in general have been systematically studied, gender differences in international collaboration have been a rare scholarly topic.

The crucial dimension in analyzing the role of female scientists in the academic enterprise is evident changes over the years. This role has changed substantially in the past 50 years (Larivière *et al*., 2013; Halevi, 2019; Huang *et al*., 2020): female scientists have increasingly been occupying high academic positions (Zippel, 2017; Madison and Fahlman, 2020) across an increasing number of disciplines (Diezmann and Grieshaber, 2019). The gender productivity, citation, and promotion gaps have been changing over time, albeit slowly, as recent overviews of gender disparities in science show (e.g., Mayer and Rathmann, 2018; Halevi, 2019; Elsevier, 2020; Huang *et al*., 2020). Specifically, Polish females constitute a substantial (43.6%), highly productive, and internationalized part of the academic profession, and Poland has a relatively high proportion of full professors compared with most Western European countries, reaching 24.1% in 2016 and continuing to increase, as the *She Figures 2018* report shows (European Commission, 2019; Kwiek, 2020b).

Females’ rising participation in academic science changes the context in which gender disparities in international research collaboration are analyzed today. New bibliometric studies increasingly apply the various gender-determination methods to authors and authorships (Halevi, 2019), and gender disparities in science are studied comprehensively (e.g., Diezmann and Grieshaber (2019), which focuses on female full professors), revealing the scale of ongoing changes. For instance, Madison and Fahlman (2020) demonstrate, for the entire population of Swedish full professors, that no bias against females occurred in attaining the rank of full professor in relation to their publication metrics. A comprehensive meta-analysis of gender effects in the peer reviews of grant proposals, with gender effect generalized over country, discipline, and publication year, indicates no evidence of any gender effects for men (Marsh *et al*., 2009, p. 1311). However, the increased participation of women in science, technology, engineering, and mathematics (STEM) disciplines is also reported to have been accompanied by an increase in gender differences regarding productivity and impact (Huang *et al*., 2020, p. 8). In general, despite ongoing transformations in their participation in academic science globally, female scientists are still reported to occupy more junior positions and to have lower salaries, to be more often in non–tenure-track and teaching-only positions, to receive less grant money, to be promoted more slowly, and to be less likely to be listed as either the first or last author on a paper, as is highlighted by the past decade of research at various levels, from institutional to national to global (see Aksnes *et al*., 2011; Larivière *et al*., 2011; Fell and König, 2016; Nielsen, 2016; Aksnes *et al*., 2019; Holman and Morandin, 2019; Maddi *et al*., 2019; Horta *et al*., 2020; Huang *et al*., 2020). Gender still predicts academic rank: women are less likely than men to hold higher ranks, and especially to attain the rank of full professor (Fox, 2020, p. 1002).

Women in science may also suffer from “biased attention” to their work (Lerchenmueller, Hoisl, & Schmallenbach, 2016) as the author’s gender is reportedly correlated with the citations received (Potthof & Zimmermann, 2017): as the proportion of women per article increases, the citations tend to decrease (as Maddi *et al*., 2019, show for economics). Moreover, the gender citation gap matters because citations are one of the chief metrics used in academia to evaluate a scholar’s performance and influence and to distribute resources, including salary (Maliniak, Powers, & Walter, 2013, p. 895), with the citation measure being increasingly used as a “reward currency in science” upon which decisions on all major aspects of an academic career are often based (Ghiasi, Mongeon, Sugimoto, & Larivière, 2018, p. 1519).

Kwiek (2018a) examined the relationships between salaries and academic behaviors and productivity across 10 European systems, and found out that being female is a very strong predictor of not being an academic top earner in Poland: female academics are, on average, highly unlikely (about one-third as likely as male academics) to belong to highly paid academics.

While men are found to produce more publications during their Ph.D. and postdoctoral years, females in these years are often diverted by marriage, starting a family, and childbearing (Halevi, 2019). In addition, the collaboration patterns and professional networks of female scientists reflect their greater focus on teaching and service compared to males. At the same time, the danger is that the objective "meritocracy-driven reliance on quantitative measures of scientific output" may, in fact, prevent female researchers from "proving their worth"—that is, from obtaining employment or moving up the academic ladder (Nielsen, 2016, p. 2057).

In addition, female scientists are reported to be less prone to leading the large-scale projects with multiple collaborators that are favored by granting agencies and increase academic visibility (Maddi *et al.*, 2019) through prestigious multiauthored international publications and their citations, thereby leading male scientists to enjoy increased collaboration, greater visibility, and more grants in the future. Bruno Latour's "credibility cycle" (Latour and Woolgar, 1986; Kwiek, 2020a, p. 3) in academic careers may be repeated faster for males than for females (each cycle leading from prestigious publications to collaborative grants to new collaborative publications to more prestige and reputation and, again, to new collaborative grants), which partially explains the comparative advantage of male scientists and their faster academic promotions. Most importantly for our study, "gender shapes patterns of international research collaboration" (Fox, 2020, p. 1003). While the role of females in science is worlds apart from that of half a century ago when classical studies were produced, the gender gaps continue and need to be analyzed for both theoretical and, perhaps especially, practical purposes.

## 2.2 *Bibliometric and Survey-Based Studies on Gender Disparities*

Bibliometric studies usually refer to international research collaboration defined as the production of internationally coauthored publications; in contrast, survey-based studies usually define international research collaboration as research conducted with international collaborators. However, both survey and bibliometric approaches are closely linked and examine related phenomena from different angles; both approaches tend to show that male and female scientists may collaborate differently (perhaps except for top performers; see Kwiek, 2018b on Polish and Kwiek 2016 on European top performers; Abramo *et al.*, 2019 on Italian top performers; and Yemini, 2019 on top performers from Denmark, Israel, and Australia). However, the evidence is inconclusive: while in some studies, female scientists tend to be less focused on international collaborations (Abramo *et al.*, 2013; Rostan *et al.*, 2014; Vabø *et al.*, 2014; Uhly *et al.*, 2015; Nielsen, 2016), others show that sex differences in international collaboration are insignificant (Larivière *et al.*, 2011; Aksnes *et al.*, 2019).

In the long run, the globalization of science as it is currently developing (including the sustained global focus on international collaborative research, large-scale research grants, the overwhelming role of top journals in academic knowledge production, and the increasing global role of productivity metrics in career progression) may present greater disadvantages for female scientists than for male scientists. Specifically, the growing importance of international research collaboration, including international mobility, in academic promotions entails comparative disadvantages for females (Zippel, 2017) who are, on average, less internationally mobile. While at earlier career stages and at a younger age, female researchers tend to be equally or more internationally mobile than male researchers, at advanced career stages and beyond the average age of 35, female researchers' flexibility to relocate internationally for more than one month decreases much more than that of their male colleagues, as a study of about 2,000 returned surveys in Germany shows (Jöns, 2011, p. 205).

In the Polish context, individual research productivity is strongly correlated with international collaboration. Polish research is characterized by two parallel processes that can be termed "internationalization accumulative advantage" and "localization accumulative disadvantage." As more international collaboration tends to imply higher publishing rates (and higher citation rates), research internationalization plays an increasingly stratifying role within the Polish academic profession, thereby leading to internationalization accumulative advantage for some scientists (Kwiek, 2019; in Europe, see Kwiek, 2015a). Increasingly, in the specific Polish context of one full decade (2010–2020) of uninterrupted higher education reforms, those who do not collaborate internationally are likely to suffer localization accumulative disadvantage in terms of resources and prestige. The male/female distinction is particularly relevant in this context, as male scientists are more internationalized—in terms of collaboration and mobility rather than merely coauthorships—in research than female scientists. "International reputation" (Frehill & Zippel, 2010, p. 50) and evidence of "international stature" in the United States (Fox, 2020, p. 1004) increasingly matter for advancement to full professorships in Poland as well. These two parallel gendered processes divide the Polish academic community—both across institutions (vertical differentiation) and across faculties within institutions (horizontal segmentation)—into highly internationalized institutions, faculties, research groups, and individual scientists versus their less internationalized, or more localized, counterparts.

In terms of survey-based studies, beyond the numerous studies on general research collaboration and gender, a few studies have focused specifically on the role of gender in international research collaboration. The findings indicate that being female is a negative predictor of international research collaboration (Rostan *et al*., 2014; Vabø *et al*., 2014; Nielsen, 2016) as the prototypical academic figure in international research collaboration is a male full professor in his mid-50s (Rostan *et al*., 2014). Vabø *et al*. (2014) found that female scientists report a lower amount of international research collaboration than do males. However, when the data are disaggregated by academic rank, the significance of the gender gap among junior faculty disappears in certain countries (i.e., the United States, Canada, South Africa, and Australia). Moreover, while male scientists are reported to be generally more involved in international research collaboration, female academics tend to be more involved in internationalization at home—for example, by teaching in a foreign language (Vabø *et al*., 2014, p. 202). Uhly *et al*. (2015, p. 14), in their cross-national comparative study of 12,959 scientists from 10 countries, report that women are significantly less likely than men to engage in international research collaboration. As they emphasize in their conclusions, "as international research collaboration impacts publication and productivity measures, which are vital for career advancement and professional recognition … gendered access to this activity has implications for broader academic stratifications" (Uhly *et al*., 2015, p. 15). Survey-based studies also reveal that being male significantly increases the odds of involvement in international research collaboration (by 69%) in 11 European countries (Kwiek, 2018c). In Fox *et al*. (2017, p. 1304), U.S. female engineers identified funding and finding international collaborators as two significant external barriers to internationalization. Women are located, disproportionately, in U.S. institutions with inadequate resources to support international research collaboration (Fox, 2020, p. 1004). While there are no significant differences in the likelihood of U.S. women versus men having a close international collaborator, the resources accessed through these relationships do show differences by gender (Melkers and Kiopa, 2010, p. 410).

### 2.3 *The Present Study*

As our study is based on bibliometric, biographical, and administrative data, we will focus briefly on one global (Larivière *et al*., 2013) and two national (Abramo *et al*., 2013; Aksnes *et al*., 2019) studies that discuss gender differences in international research collaboration from similar perspectives. Larivière and colleagues analyzed 5.48 million research papers and review articles with more than 27 million authorships in the 2008–2012 period from the Web of Science database. For the 50 most productive countries in their analysis, they found that female collaborations are more domestically oriented than collaborations of males from the same country. The study, due to its scale and global character, was unable to include academic positions, age, or disciplines in its analysis, all known from previous literature to have a powerful impact on international collaboration rate.

Bibliometric research on gender disparity in international collaboration at a national level has been conducted in Norway and Italy. The general conclusion for Norway was that the propensity to collaborate in international research was similar for both male and female scientists (Aksnes *et al*., 2019); however, for Italy, this propensity is higher for male scientists across the entire population (Abramo *et al*., 2013), being similar for male and female top performers (Abramo *et al*., 2019). Both Italian and Norwegian studies have addressed the gap in research on gender differences both in research collaboration in general and international research collaboration in particular by taking the individual scientist—rather than the individual publication—as the unit of analysis. In the case of all Italian scientists, Abramo *et al*. (2013), using almost 200,000 Web of Science publications in the 2006–2010 period by 36,211 productive scientists, showed that female scientists are more likely to collaborate domestically both intramurally and extramurally but are less likely to engage in international collaboration. The methodology used in their study avoids distortion by outliers—that is, by cases of highly productive and highly internationalized scientists whose extensive publications distort aggregate index values. The same approach is adopted in our paper.

In Norway, Aksnes *et al*. (2019) used the Cristin bibliographic database (Norwegian Science Index of all peer-reviewed publications) and counted all individuals equally as single units, regardless of their productivity, limiting the effect of the outliers. They used a database that has complete coverage of all Norwegian peer-reviewed and scholarly publication output, including books, edited volumes, and conference series. However, the researchers included ($N = 5{,}554$, with more than 43,000 publications from the 2015–2017 period) come from the four largest universities only. Gender differences were analyzed by field, academic position, and publication productivity, and scientific discipline emerged as the most important determinant of international research collaboration. However, gender differences were not statistically significant.

The difference between these three studies and this paper is as follows. While not global in nature, this paper, using a combination of biographical, administrative and bibliometric data from a single system, analyzes all Scopus-indexed articles published in the 2009–2018 period by all internationally visible Polish academic scientists. All scientists have their gender, academic position, and age clearly defined due to the nature of the new data set produced by the authors; additionally, all scientists are ascribed to ASJC disciplines. While Larivière *et al*. (2013) did not use disciplines, academic positions, or age, Abramo *et al*. (2013) focused on disciplines and all major collaboration types, and Aksnes *et al*. (2019) focused on fields, disciplines, academic positions, and collaboration intensity, our study combines all the above dimensions and adds biological age (and age groups of young, middle-aged, and older scientists) as a dependent variable since age is certainly "one of the most apparent personal factors one might expect to have an effect on collaborations" (Bozeman *et al*., 2012, p. 7).

Few studies combine age, academic position, and international research collaboration—as we do in our paper—because only a few data sets combine administrative and biographical data at the individual level, on the one hand, and publication and citation data, on the other. These combinations are studied mostly at the level of individual institutions where data availability is higher; large-scale studies at the national level require either sophisticated data set mergers or access to comprehensive registry-based national databases. Given the policy challenges posed by the progressive aging of European scientists, data-driven studies of national populations of scientists, as well as their propensity to collaborate by age (and age group), seem particularly useful.

In principle, collaboration can be examined by age, academic cohort, and period, and the respective age, cohort, and period effects need to be carefully distinguished. However, in practical terms, "except under conditions that hardly ever exist, a definitive separation of age, period, and cohort effects is not just difficult, but impossible" (Glenn, 2005, p. vii). This research is cross-sectional (i.e., scientists are not followed over time as in longitudinal studies); therefore, age and cohort effects are intermingled. We cannot observe individuals born at different dates at several points in time (Hall *et al*., 2007, pp. 159–161). Differences by age shown in this paper may or may not be age effects because Polish scientists of different ages are members of different cohorts and "may have been shaped by different formative experiences and influences," with differences between them possibly being cohort effects (Glenn, 2005, p. 3). In this paper, as we are unable to study change over time, we present a snapshot of a single decade. All we can learn from our research pertains to male and female scientists of varying ages publishing in the period 2009–2018. Although "cohort matters" and careers of scientists are clearly affected by "events occurring at the time their cohort graduates" Stephan, ,2012, pp. 174–175), cohort analysis *par excellence* cannot be conducted based on the data set at our disposal. Belonging to a specific historical generation can have an influence on individual collaboration patterns; individual opportunities to engage in international collaboration differ by period (Rostan *et al*., 2014, p. 125), which is evident, for instance, when comparing cohorts prior to the collapse of the communist regime in Poland in 1989 and afterward, when the Polish science system was gradually opening to global collaboration in research (see different "academic generations" as examined in Kwiek, 2015b). Following Kyvik and Aksnes (2015) and Rørstadt and Aksnes (2015), who suggested that there has been a general change in the norms of academic publishing behavior in the case of Norwegian scientists over recent decades, in the Polish case, we can assume that younger cohorts not only tend to publish more today than younger cohorts a decade or two decades ago but tend to publish more in highly prestigious journals and more often in international collaboration.

As they clearly demonstrated for the youngest Norwegian age cohort of the 1989–1991 period (Kyvik and Aksnes, 2015, p. 1448), certain generations excel in international collaboration over time and as they age—younger and older Polish academics are textbook examples of this. Career opportunities and academic norms differed significantly for those entering the academic profession prior to 1989 and those who came after; the same is applicable to those who entered the Polish profession before and after the reforms of the 2010s.

Gender disparity in Polish science has rarely been studied, and gender collaboration patterns have not been examined. Kosmulski (2015) analyzed the productivity and impact of male and female scientists in the period 1975–2014 based on a limited set of authors bearing one of the 26 most popular "–ski" or "–cki" names, showing that male scientists generally have higher productivity and impact, except for in biochemistry, where the genders are almost equal. Siemienska (2007) based her research on two small-scale surveys of full professors and young academics and revealed that cultural capital (measured as the level of parents' education) was particularly important for the research productivity of females. As measured by a proxy of internationally coauthored publications, Poland had the lowest level of research internationalization in the European Union in 2019 (36.1% based on Scopus data, articles only).

## 3. Data and Methods

### 3.1 *The Data Set*

Two large databases were merged: Database I was an official national administrative and biographical register of all Polish scientists; Database II was the Scopus publication and citation database. Database I comprised 99,935 scientists employed in the Polish science sector as of November 21, 2017. Only scientists with at least a doctoral degree (70,272) and employed in the higher education sector were selected for further analysis (54,448 scientists). The data used were both demographic and professional, with each scientist identified by a unique ID. Database II, the original Scopus database, included 169,775 names from all 85 universities with internationally visible publications within the decade. Authors in Database II were defined by their institutional affiliations, their Scopus documents, and individual Scopus IDs.

The key procedure was to appropriately identify authors with their different individual IDs in the two databases and to provide them with a new ID in the integrated "Polish Science Observatory" database. Probabilistic methods of data integration were used (as defined in Fellegi and Sunter 1969 and Enamorado *et al.*, 2019). Separately within each of the 85 universities, the first name and last name records of each record in Database I were compared with each of the records in Database II using the Jaro–Winkler string distance (see Jaro, 1989; Winkler, 1990). Next, using an expectation maximization algorithm (Enamorado *et al.*, 2019), the posterior probability that a given pair of records belongs to the same unit was estimated. If the probability was greater than 0.85, the pair was considered to be part of the same unit (Harron *et al.*, 2017). The computation was made using the fastLink R package (version 0.6.0). An integrated database obtained in accordance with the above procedures and used in our research finally included 37,081 records, which were referring to 32,937 unique authors.[1] Database I contained biographical and professional career information on all authors affiliated with the 85 largest Polish universities in the 2009–2018 reference period. Database II contained metadata on 377,886 publications with up to 100 authors. From among the 377,886 publications in the original Database II, 230,007 were written by the authors included in Database I. Subsequently, only articles written in journals were selected for further analysis, all other publication types being omitted. The number of papers in the database was therefore reduced to 158,743 articles and the number of unique authors was reduced to 25,463. Approximately half of the Polish scientists from the higher education sector (53.4%) did not publish a paper indexed in the Scopus database in the reference period—which is in line with previous findings regarding the distribution of Polish publications—with the overwhelming majority of publications belonging to national publication outlets.

### 3.2 *Methods*

Every Polish scientist represented in our integrated database was ascribed to one of 334 ASJC disciplines at the four-digit level and one of 27 ASJC disciplines at the two-digit level (following Abramo *et al.*, 2020, who defined in their study the dominant Web of Science subject category for each Italian and Norwegian professor). In the ASJC system used, a given paper can have one or multiple disciplinary classifications.[2] The dominant ASJC for each scientist was determined (the most frequently occurring value). In the case when no single mode occurred, the dominant ASJC was randomly selected. Consequently, we had Polish scientists defined by their gender and ASJC discipline, along with all their publications from the decade studied. We also had a proportion of female scientists in every ASJC discipline. Furthermore, three disciplines were omitted from analysis as they did not meet an arbitrary minimum threshold of 50 scientists per discipline (GEN, NEURO, and NURS).

Having an individual scientist as the unit of analysis, we calculated the proportion of internationally coauthored articles among collaborative articles within the individual publication portfolio of every Polish scientist in the sample. Gender determination of names was not necessary: the administrative and biographical Database I contained gender information for all the observations. The integrated data set of Polish scientists with their administrative, biographical, publication, and citation data used in this research (termed "The Polish Science Observatory") is maintained and periodically updated by the two authors as part of ongoing research programs in the Center for Public Policy Studies.

### 3.3 *The Sample*

The structure of the sample ($N = 25{,}463$) is presented in Table 1: approximately half of the scientists are in the 36–50 years age bracket (51.5%), and over half of them are assistant professors (56.0%).

**Table 1.** Structure of the Sample, all Polish Internationally Visible University Professors, by Gender, Age Group, Academic Position, and Discipline, Presented with Column and Row Percentages.

| | | Female | | | Male | | | Total | | |
|---|---|---|---|---|---|---|---|---|---|---|
| | | *N* | % col | % row | *N* | % col | % row | *N* | % col | % row |
| Age group | 26–30 | 245 | 2.3 | 46.8 | 279 | 1.9 | 53.2 | 524 | 2.1 | 100.0 |
| | 31–35 | 1 653 | 15.6 | 49.8 | 1 664 | 11.2 | 50.2 | 3 317 | 13.0 | 100.0 |
| | 36–40 | 2 148 | 20.3 | 47.1 | 2 411 | 16.2 | 52.9 | 4 559 | 17.9 | 100.0 |
| | 41–45 | 2 272 | 21.5 | 45.7 | 2 696 | 18.1 | 54.3 | 4 968 | 19.5 | 100.0 |
| | 46–50 | 1 569 | 14.8 | 43.8 | 2 013 | 13.5 | 56.2 | 3 582 | 14.1 | 100.0 |
| | 51–55 | 993 | 9.4 | 40.3 | 1 471 | 9.9 | 59.7 | 2 464 | 9.7 | 100.0 |
| | 56–60 | 671 | 6.3 | 37.7 | 1 108 | 7.4 | 62.3 | 1 779 | 7.0 | 100.0 |
| | 61–65 | 563 | 5.3 | 26.7 | 1 548 | 10.4 | 73.3 | 2 111 | 8.3 | 100.0 |
| | 66–70 | 417 | 3.9 | 23.0 | 1 396 | 9.4 | 77.0 | 1 813 | 7.1 | 100.0 |
| | 71+ | 46 | 0.4 | 13.3 | 300 | 2.0 | 86.7 | 346 | 1.4 | 100.0 |
| | Total | 10 577 | 100.0 | 41.5 | 14 886 | 100.0 | 58.5 | 25 463 | 100.0 | 100.0 |
| Academic position | Assistant professor | 6 851 | 64.8 | 48.0 | 7 420 | 49.8 | 52.0 | 14 271 | 56.0 | 100.0 |
| | Associate professor | 2 822 | 26.7 | 38.0 | 4 596 | 30.9 | 62.0 | 7 418 | 29.1 | 100.0 |
| | Full professor | 904 | 8.5 | 24.0 | 2 870 | 19.3 | 76.0 | 3 774 | 14.8 | 100.0 |
| | Total | 10 577 | 100.0 | 41.5 | 14 886 | 100.0 | 58.5 | 25 463 | 100.0 | 100.0 |
| Discipline (ASJC) – STEM | | | | | | | | | | |
| | AGRI | 1 444 | 13.7 | 53.4 | 1 258 | 8.5 | 46.6 | 2 702 | 10.6 | 100.0 |
| | BIO | 1 068 | 10.1 | 60.0 | 712 | 4.8 | 40.0 | 1 780 | 7.0 | 100.0 |
| | CHEM | 756 | 7.1 | 51.3 | 719 | 4.8 | 48.7 | 1 475 | 5.8 | 100.0 |
| | CHEMENG | 185 | 1.7 | 38.5 | 296 | 2.0 | 61.5 | 481 | 1.9 | 100.0 |
| | COMP | 170 | 1.6 | 16.5 | 860 | 5.8 | 83.5 | 1 030 | 4.0 | 100.0 |
| | DEC | 24 | 0.2 | 44.4 | 30 | 0.2 | 55.6 | 54 | 0.2 | 100.0 |
| | EARTH | 385 | 3.6 | 33.4 | 769 | 5.2 | 66.6 | 1 154 | 4.5 | 100.0 |
| | ENER | 82 | 0.8 | 27.8 | 213 | 1.4 | 72.2 | 295 | 1.2 | 100.0 |
| | ENG | 501 | 4.7 | 14.9 | 2 857 | 19.2 | 85.1 | 3 358 | 13.2 | 100.0 |
| | ENVIR | 848 | 8.0 | 50.5 | 832 | 5.6 | 49.5 | 1 680 | 6.6 | 100.0 |
| | IMMU | 90 | 0.9 | 75.6 | 29 | 0.2 | 24.4 | 119 | 0.5 | 100.0 |
| | MATER | 495 | 4.7 | 33.9 | 967 | 6.5 | 66.1 | 1 462 | 5.7 | 100.0 |

(*Continued*)

**Table 1.** *Continued.*

| | | Female | | | Male | | | Total | | |
|---|---|---|---|---|---|---|---|---|---|---|
| | | *N* | % col | % row | *N* | % col | % row | *N* | % col | % row |
| | MATH | 259 | 2.4 | 25.2 | 767 | 5.2 | 74.8 | 1 026 | 4.0 | 100.0 |
| | PHARM | 169 | 1.6 | 66.5 | 85 | 0.6 | 33.5 | 254 | 1.0 | 100.0 |
| | PHYS | 182 | 1.7 | 16.6 | 916 | 6.2 | 83.4 | 1 098 | 4.3 | 100.0 |
| Discipline (ASJC) – non-STEM | BUS | 372 | 3.5 | 52.1 | 342 | 2.3 | 47.9 | 714 | 2.8 | 100.0 |
| | DENT | 57 | 0.5 | 76.0 | 18 | 0.1 | 24.0 | 75 | 0.3 | 100.0 |
| | ECON | 186 | 1.8 | 49.1 | 193 | 1.3 | 50.9 | 379 | 1.5 | 100.0 |
| | HEALTH | 23 | 0.2 | 34.3 | 44 | 0.3 | 65.7 | 67 | 0.3 | 100.0 |
| | HUM | 527 | 5.0 | 49.8 | 531 | 3.6 | 50.2 | 1 058 | 4.2 | 100.0 |
| | MED | 1 920 | 18.2 | 53.7 | 1 654 | 11.1 | 46.3 | 3 574 | 14.0 | 100.0 |
| | PSYCH | 194 | 1.8 | 63.8 | 110 | 0.7 | 36.2 | 304 | 1.2 | 100.0 |
| | SOC | 494 | 4.7 | 49.8 | 498 | 3.3 | 50.2 | 992 | 3.9 | 100.0 |
| | VET | 146 | 1.4 | 44.0 | 186 | 1.2 | 56.0 | 332 | 1.3 | 100.0 |
| | Total | 10,577 | 100.0 | 41.5 | 14,886 | 100.0 | 58.5 | 25,463 | 100.0 | 100.0 |

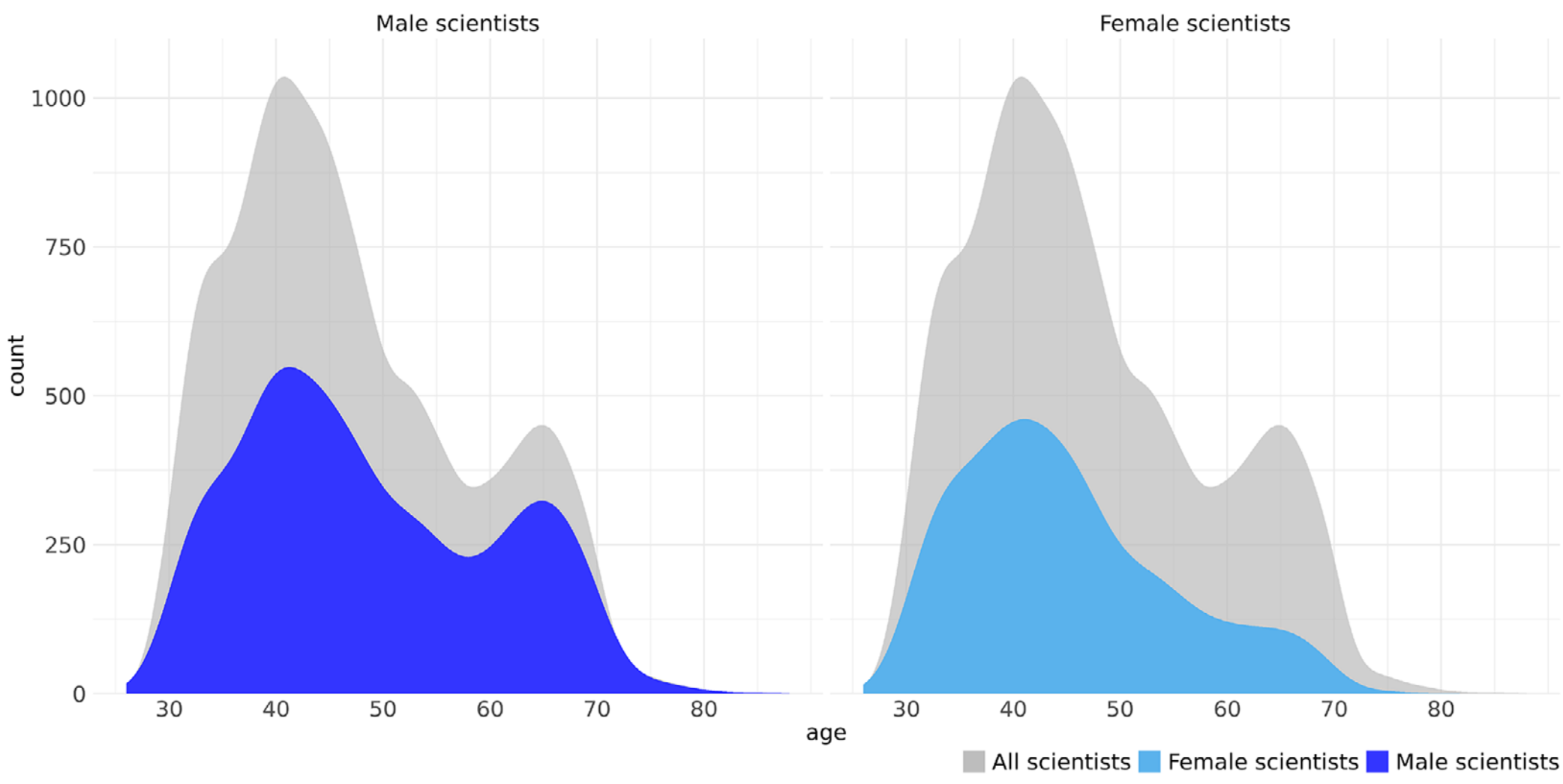

**Figure 1.** Age Structure of the Sample, All Polish Internationally Visible University Professors ($N = 25{,}463$), by Gender. All University Professors in Gray. [Colour figure can be viewed at wileyonlinelibrary.com]

Column percentages enable the analysis of the gender distribution of the Polish academic profession by age groups, academic positions, and disciplines, while row percentages enable the analysis of how male and female scientists are distributed according to a given age group, academic position, and discipline. Table A1 in the Appendix shows age distributions for each academic position from a gender perspective. The three largest disciplines represented in the sample are agricultural and biological sciences, engineering, and medicine (AGR, ENG, and MED), representing over one-third of the scientists (37.8%). The list of gender-balanced disciplines goes beyond the social sciences and humanities (to include also business, economics, agricultural and biological sciences, medicine, chemistry and biochemistry, genetics, and psychology). Of the 24 ASJC Scopus disciplines studied in this paper, female representation reaches at least 50% in 10 of them.

Female participation in the academic profession decreases with age: while female scientists represent approximately half of all scientists aged 31–35, they represent only about a quarter of all scientists aged 61–65 years (49.8% and 26.7%, respectively). Female scientists are also clustered in lower academic positions: while females constitute about half of all assistant professors, they represent only about a quarter of full professors (48% and 24%, respectively, levels comparable to those in many other countries; for Sweden, see Madison and Fahlman, 2020, and for global overviews, see Larivière *et al*., 2013; Halevi, 2019; and Diezmann and Grieshaber, 2019). Polish assistant professors under 45 (our entire sample includes scientists with doctorates only) have an almost equal gender distribution. The older professors (aged 41–55 years) with a habilitation degree (a second, postdoctoral degree) are already dominated by male scientists (who represent approximately 60% of associate professors). In the case of full professors, the number of males is at least three times that of females (see Table A1 in Appendix) for every age group for both young full professors (aged 41–45) and the oldest ones (aged 61–65). All assistant professors as defined in this paper hold doctoral degrees, all associate professors hold habilitations, and all full professors hold full professorships.

The age structure by gender of the sample is presented in Figure 1. Our sample contains only scientists who had at least a single article in the Scopus database in the period 2009–2018 and, therefore, it includes all internationally visible Polish academic scientists (on highly skewed research productivity of Polish scientists, see Kwiek, 2018b). The differentiated proportions of female scientists can also be examined by academic discipline. Female scientists are severely underrepresented in computer science (COMP 16.5%), engineering (ENG 14.9%), physics and astronomy (PHYS 16.6%), and mathematics (MATHS 25.2%). In arts and humanities (HUM) and social sciences (SOC), the distribution of scientists by gender is practically equal (49.8%).

International research collaboration (defined as the occurrence of an article with at least two authors, of which at least one has a non-Polish institutional affiliation) is examined in this paper in the context of three other collaboration types: collaboration in general (defined as the occurrence of an article with at least two authors), national collaboration (article with at least two authors with two different Polish affiliations), and institutional collaboration (article with at least two authors with the same Polish affiliation). An article published in an international collaboration can also be counted as an article published in national collaboration (if it has authors with at least two different Polish affiliations) and institutional collaboration (if it has authors with at least two of the same Polish affiliations). An article published in national collaboration can also be counted as an article published in institutional collaboration (if at least two authors have the same Polish institutional collaboration), following traditional distinctions between collaboration types.

## 4. Results

### 4.1 *International Research Collaboration: Distribution of Output by Gender*

Of all internationally visible Polish university professors (25,463), slightly less than half (11,854 or 46.6%) collaborate internationally: 7,057 males and 4,797 females. The share of females involved in international collaboration is only slightly lower (45.35%) than the share of males (47.41%). Their involvement in international research collaboration is defined as having at least a single article written in international collaboration in the decade examined. However, the scale of this involvement is low, as shown in Figure 2. The distribution of the number of articles published in international research collaboration is highly skewed, with a long tail on the right, indicating extreme inequality. About one-third of scientists authored only a single article, and about a half authored at least two articles (33.4% and 50.2%, respectively; 11.9% authored at least 10 articles, whereas 3.8% authored at least 20 articles). Figure 3 shows the gender difference in international collaboration: females fare consistently slightly worse for every threshold in international collaboration, starting with only one article (68.8% of males vs. 63.4% of females published more than a single article). Moreover, the rates of decrease for females were steeper; for instance, the percentage of males with more than 10, 15, and 20 articles is twice as high as the same rate for females (14.5% vs. 8.1%, 8.3% vs. 4.1%, and 4.9% vs. 2.1%). For every number of articles published, the share of males with at least this number is higher for males than for females (compare the two lines).

Table 2 summarizes the distribution statistics: specifically, the respective 5% trimmed means in the case of males and females are significantly smaller than the arithmetic mean (calculated on the basis of all observations) and lie outside their 95% confidence intervals. This means that there are observations with values significantly larger than typical ones. Very high variability in the number of articles is demonstrated by the coefficient of variance (the ratio of the standard deviation and arithmetic mean). In both cases, it is substantially greater than 1—while a distribution with very high variability is considered to be one for which this factor is greater than 0.5. The asymmetry of the distribution of the number of articles for both males and females is extreme and right-tailed, as evidenced by the very high positive values of the asymmetry coefficients (9.14 and 5.67, respectively).

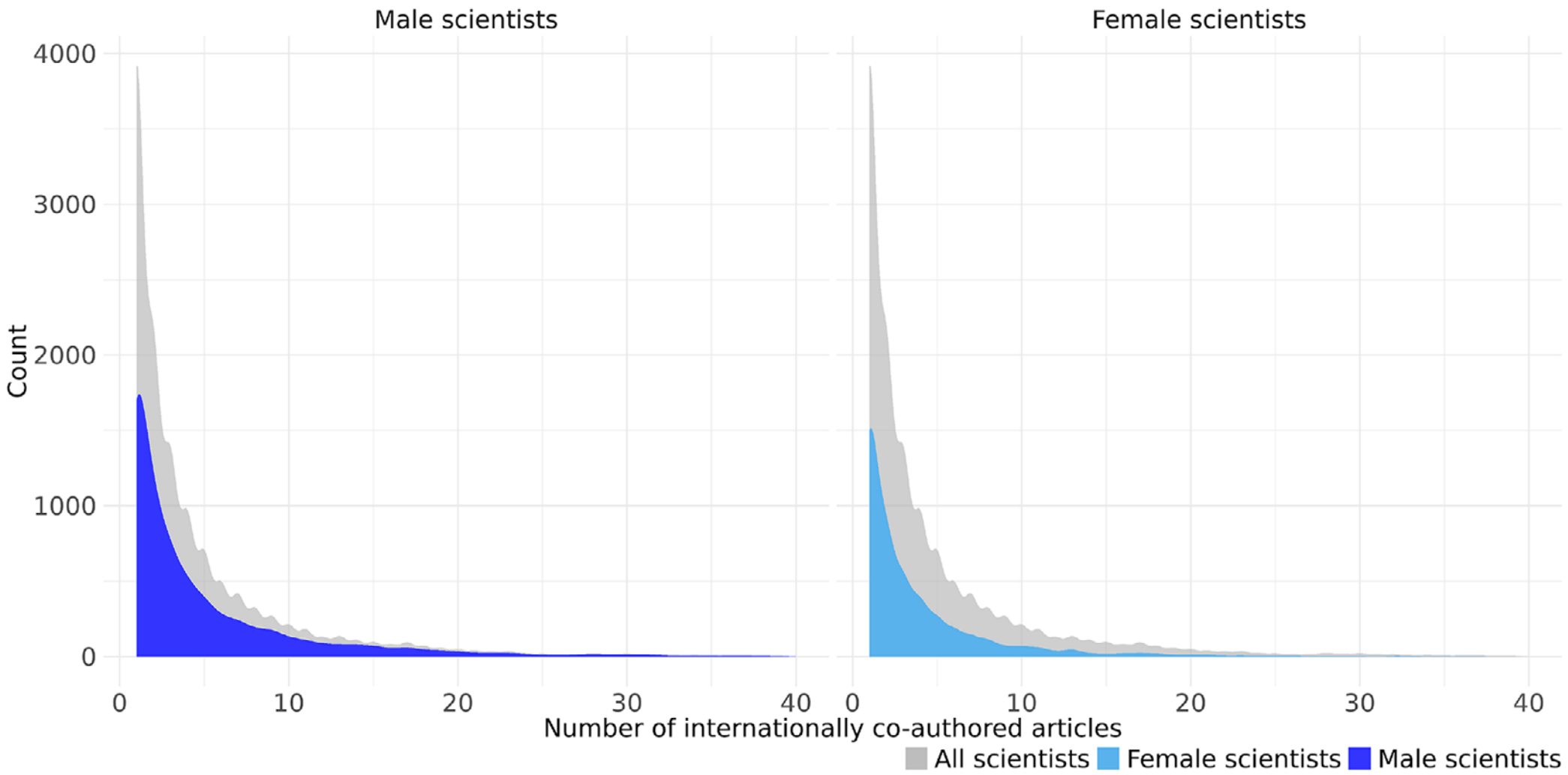

**Figure 2.** The Distribution of the Number of Articles Published in International Research Collaboration by Gender, 2009–2018 Combined ($N = 11{,}854$, Cutoff Point of 40 Articles Used). [Colour figure can be viewed at wileyonlinelibrary.com]

## 4.2 *Four Collaboration Types and Three Collaboration Intensity Levels*

Three approaches to measuring international research collaboration were tested in this paper: a minimum threshold approach (one article only), an over 50% approach, and an over 75% approach, representing a low, medium, and high collaboration intensity. This intensity is measured at individual level in scientists' individual publication portfolios (for instance, an author has a total of five collaborative publications, including three published in international collaboration; thus the author represents both low and medium collaboration intensity but not high intensity, as the share of internationally co-authored publications in this case is 60%). Distributions by gender and dominant discipline were examined for each approach.

Further, we examined the gender disparity for each of the four collaborative types (general, international, national, and institutional) according to the three levels of collaboration intensity. In the case of general collaboration (any collaboration type, a superset of all other collaboration types), the gender disparity emerges as differentiated by discipline and collaboration intensity. While at low levels of intensity in general collaboration, gender disparities are negligible, they increase with intensity. The same pattern is observed for national, institutional, and international collaboration (for example, in the case of institutional collaboration—a definitely dominating collaboration type in Poland—gender differences increase with collaboration intensity in the following manner: from 83.1% vs. 82.6% at a low intensity level to 67.1% vs. 62.6% at a medium intensity level to 53.6% vs. 47.2% at a high intensity level). Still, they are not strikingly different, as could be expected based on previous literature.

In the case of high-intensity collaborations for all disciplines combined, collaboration rates by gender are slightly higher for females in general collaboration, institutional collaboration, and national collaboration (83.2% vs. 80.6%, 53.6% vs. 47.2%, and 4.5% vs. 3.9%, respectively, see Total in Table 3). Male scientists exhibit slightly higher collaboration rate only for the most demanding and most expensive collaboration type: international collaboration (4.1% and 5.2%).

However, the data analysis for all disciplines combined does not tell the entire story of gender disparity. There is a fascinating cross-disciplinary gender disparity in all four collaboration types. In general, in national and institutional collaboration, there are specific disciplines in which male scientists exhibit higher collaboration propensity, against the picture at the aggregated level of all disciplines; and, in a similar vein, in international collaboration, there are specific disciplines in which female scientists exhibit higher collaboration propensity. Figure 4 presents gender differences by collaboration type (four panels) and discipline in greater detail: results above zero indicate a female advantage in a given discipline and results below zero indicate a male advantage. The differences by discipline are presented for high collaboration intensity only.

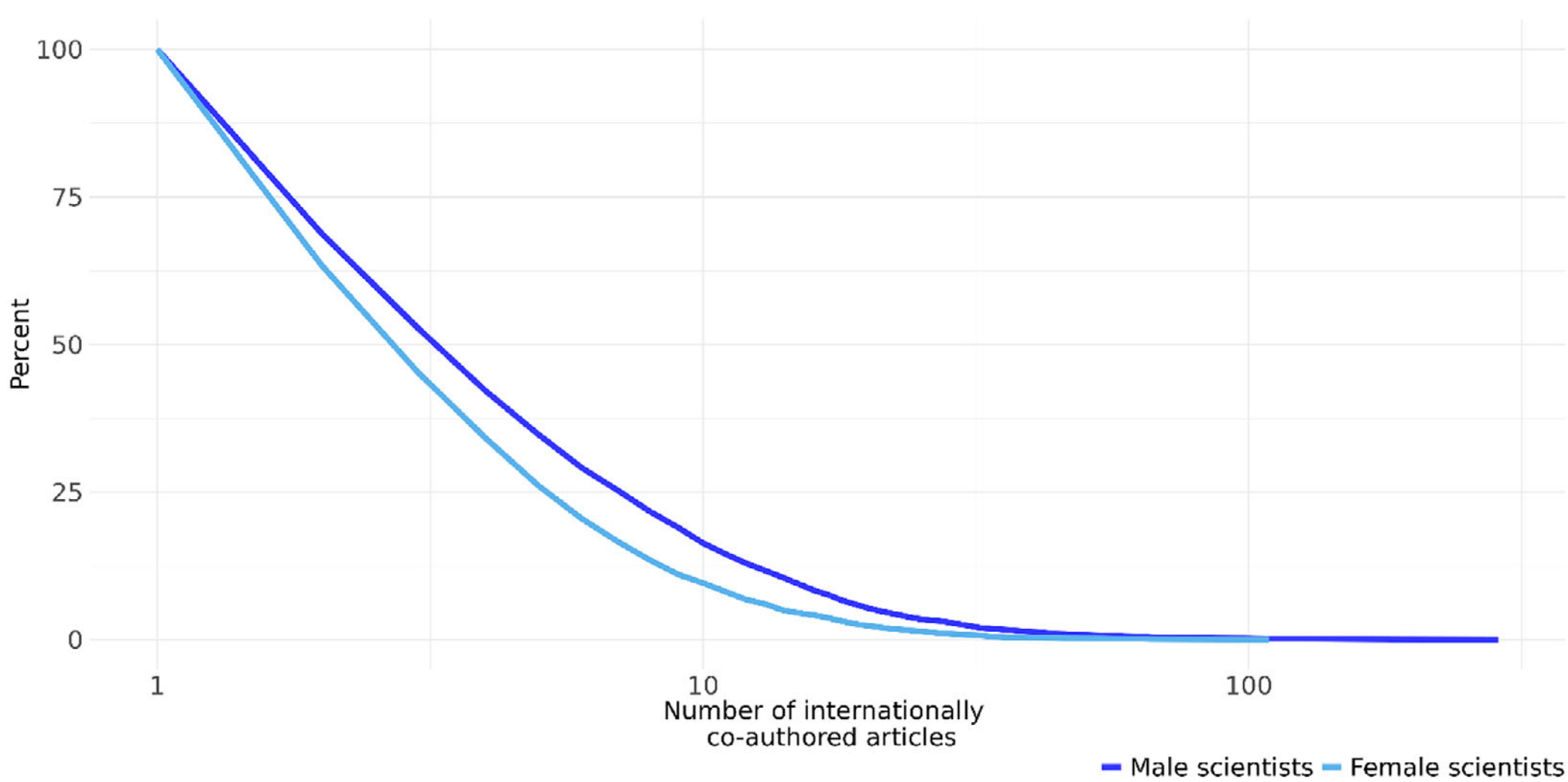

**Figure 3.** The Reverse Cumulative Distribution Function of the Log Number of Articles Published in International Research Collaboration by Gender, 2009–2018 Combined ($N = 11{,}854$). [Colour figure can be viewed at wileyonlinelibrary.com]

In the case of general collaboration, the propensity of male scientists to collaborate is higher in 13 out of 24 disciplines (see the top left panel in Figure 4). Among the three largest disciplines (AGRI, ENG, and MED), female scientists show higher propensity in only one—AGRI. In contrast to Abramo *et al*. (2013), who found higher propensity to general collaboration among Italian females for almost all disciplines, the propensity to collaborate in general is higher for Polish male scientists for over half of all disciplines, with the highest percentage difference for ECON—reaching 6.0 percentage points. For national collaboration (the bottom left panel), the male advantage was found for 10 disciplines, with HUM exhibiting no gender difference. The female advantage reached higher levels for HEALTH and ENERGY, two small disciplines. For institutional collaboration (the top right panel), the male advantage was found for 13 disciplines.

The propensity to collaborate internationally deserves a separate treatment. Overall, international collaboration rates are low for all intensity levels (a finding that is in line with findings for Poland at the highly aggregated level: 36.1% of Scopus-indexed articles in 2019 were written in international collaboration, which is the lowest rate among the European Union member states).

First, we will study the low level of international collaboration intensity (a minimum threshold approach). In our sample, less than a half of female (45.4%) and male (47.4%) scientists had at least one paper published in international collaboration (total 46.6%; see Table A2 in the Appendix). In the three largest disciplines, the rate of international collaboration was higher for male scientists by as much as 8.3 p.p. for MED and 4.8 p.p. for ENG (for AGRI, it is 2.6 p.p.). The largest male advantage is noted for disciplines such as PSYCH, PHARM, ENER, and BIO (10.8–17.4 p.p.). For HUM, SOC, and ECON, the three disciplines with the lowest rates of international collaboration, the male advantage is notable. HUM definitely has the lowest rate: only 8.3% female scientists and 11.9% male scientists have published at least one article in international coauthorship in the decade studied. At this low level of collaboration intensity, female advantage occurs in only four disciplines: two medium-sized (BUS and CHEMENG) and two small (DEC and DENT) disciplines. There are five disciplines in which international collaboration reaches the highest levels (in the range of 60%–70%): large disciplines such as BIO, CHEM, MATER, and PHYS as well as the small discipline IMMU. Male scientists are highly likely to collaborate internationally at this low level of intensity in CHEM (75%), PHYS (74.8%), and BIO (71.2%).

**Table 2.** The Number of Articles Distribution Statistics.

| Statistic | | Male | Female |
|---|---|---|---|
| Mean | | 5.95 | 4.12 |
| 95% confidence interval | Lower bound | 5.70 | 3.95 |
| for mean | Upper bound | 6.21 | 4.28 |
| 5% trimmed mean | | 4.42 | 3.23 |
| Median | | 3.00 | 2.00 |
| Variance | | 116.791 | 33.959 |
| Std. deviation | | 10.807 | 5.827 |
| Minimum | | 1 | 1 |
| Maximum | | 286 | 109 |
| Range | | 285 | 108 |
| Interquartile range | | 6 | 4 |
| Skewness | | 9.135 | 5.668 |
| Kurtosis | | 147.753 | 57.452 |

Second, at a medium level of collaboration intensity (>50% articles published in international collaboration in the scientist's individual publication portfolio for 2009–2018), male advantage is overwhelming, with only four disciplines in which there is female advantage (again BUS, CHEMENG, DEC, and IMMU). However, as expected, collaboration rates dropped drastically compared to the low level of collaboration intensity to an average of 8.4% for female scientists and 10.8% for male scientists. Only in one discipline, at least a quarter of both male and female scientists attained this medium intensity level (PHYS: 31.0% and 25.3%, respectively) and only in several disciplines, at least 15% of scientists attained this level of intensity (females in BUS, DEC, and MATH; males in PSYCH).

And third, at the highest level of intensity (>75% articles published in international collaboration in the scientist's individual publication portfolio for 2009–2018), international collaboration rates dropped by half to 4.1% for females scientists and 5.2% for male scientists (Table A2 in the Appendix). For international collaboration, the female advantage was noted for 9 disciplines and the male advantage was noted for 13 disciplines, with CHEMENG exhibiting no gender difference (the bottom right panel in Figure 4). For the three largest disciplines, the gender disparity is marginal. The largest male advantage was noted for MATH, PSYCH, and PHYS but the gender disparity is the largest for the two very small

**Table 3.** Percentage Differences in High-Intensity Collaborations (>75% Articles Published in the Four Collaboration Types in the Scientist's Individual Publication Portfolio for 2009–2018), All Polish Internationally Visible University Professors, by Collaboration Type, Gender, and ASJC Discipline (in %). Non-STEM Disciplines begin with BUS.

| | | Scientists: total | | Scientists: general collaboration | | | Scientists: international collaboration | | | Scientists: national collaboration | | | Scientists: institutional collaboration | | |
|---|---|---|---|---|---|---|---|---|---|---|---|---|---|---|---|
| | | *N* | % col | *N* | % col | % row | *N* | % col | % row | *N* | % col | % row | *N* | % col | % row |
| AGRI | F | 1444 | 53.4 | 1361 | 53.8 | 94.3 | 44 | 59.5 | 3 | 52 | 44.8 | 3.6 | 896 | 55.2 | 62 |
| | M | 1258 | 46.6 | 1170 | 46.2 | 93 | 30 | 40.5 | 2.4 | 64 | 55.2 | 5.1 | 727 | 44.8 | 57.8 |
| BIO | F | 1068 | 60 | 1053 | 60.3 | 98.6 | 48 | 53.3 | 4.5 | 59 | 53.2 | 5.5 | 782 | 61.6 | 73.2 |
| | M | 712 | 40 | 693 | 39.7 | 97.3 | 42 | 46.7 | 5.9 | 52 | 46.8 | 7.3 | 487 | 38.4 | 68.4 |
| CHEM | F | 756 | 51.3 | 723 | 51.1 | 95.6 | 34 | 43.6 | 4.5 | 29 | 56.9 | 3.8 | 449 | 53.5 | 59.4 |
| | M | 719 | 48.7 | 691 | 48.9 | 96.1 | 44 | 56.4 | 6.1 | 22 | 43.1 | 3.1 | 390 | 46.5 | 54.2 |
| CHEM | F | 185 | 38.5 | 174 | 41.4 | 94.1 | 5 | 38.5 | 2.7 | 12 | 48 | 6.5 | 120 | 43.8 | 64.9 |
| ENG | M | 296 | 61.5 | 246 | 58.6 | 83.1 | 8 | 61.5 | 2.7 | 13 | 52 | 4.4 | 154 | 56.2 | 52 |
| COMP | F | 170 | 16.5 | 132 | 16.5 | 77.6 | 13 | 17.3 | 7.6 | 8 | 22.9 | 4.7 | 55 | 14.3 | 32.4 |
| | M | 860 | 83.5 | 669 | 83.5 | 77.8 | 62 | 82.7 | 7.2 | 27 | 77.1 | 3.1 | 330 | 85.7 | 38.4 |
| DEC | F | 24 | 44.4 | 15 | 53.6 | 62.5 | 4 | 80 | 16.7 | 1 | 50 | 4.2 | 4 | 36.4 | 16.7 |
| | M | 30 | 55.6 | 13 | 46.4 | 43.3 | 1 | 20 | 3.3 | 1 | 50 | 3.3 | 7 | 63.6 | 23.3 |
| EARTH | F | 385 | 33.4 | 272 | 32.2 | 70.6 | 28 | 43.1 | 7.3 | 10 | 21.7 | 2.6 | 125 | 30.8 | 32.5 |
| | M | 769 | 66.6 | 573 | 67.8 | 74.5 | 37 | 56.9 | 4.8 | 36 | 78.3 | 4.7 | 281 | 69.2 | 36.5 |
| ENER | F | 82 | 27.8 | 65 | 27.5 | 79.3 | 4 | 25 | 4.9 | 14 | 38.9 | 17.1 | 36 | 24.8 | 43.9 |
| | M | 213 | 72.2 | 171 | 72.5 | 80.3 | 12 | 75 | 5.6 | 22 | 61.1 | 10.3 | 109 | 75.2 | 51.2 |
| ENG | F | 501 | 14.9 | 377 | 14.7 | 75.2 | 10 | 10.9 | 2 | 25 | 26.6 | 5 | 197 | 13.7 | 39.3 |
| | M | 2857 | 85.1 | 2189 | 85.3 | 76.6 | 82 | 89.1 | 2.9 | 69 | 73.4 | 2.4 | 1242 | 86.3 | 43.5 |
| ENVIR | F | 848 | 50.5 | 749 | 49.9 | 88.3 | 16 | 38.1 | 1.9 | 33 | 50 | 3.9 | 501 | 52.1 | 59.1 |
| | M | 832 | 49.5 | 753 | 50.1 | 90.5 | 26 | 61.9 | 3.1 | 33 | 50 | 4 | 460 | 47.9 | 55.3 |
| IMMU | F | 90 | 75.6 | 86 | 74.8 | 95.6 | 9 | 90 | 10 | 3 | 75 | 3.3 | 55 | 75.3 | 61.1 |
| | M | 29 | 24.4 | 29 | 25.2 | 100 | 1 | 10 | 3.4 | 1 | 25 | 3.4 | 18 | 24.7 | 62.1 |
| MATER | F | 495 | 33.9 | 464 | 34 | 93.7 | 29 | 35.8 | 5.9 | 23 | 48.9 | 4.6 | 270 | 32.9 | 54.5 |
| | M | 967 | 66.1 | 901 | 66 | 93.2 | 52 | 64.2 | 5.4 | 24 | 51.1 | 2.5 | 551 | 67.1 | 57 |
| MATH | F | 259 | 25.2 | 154 | 26 | 59.5 | 13 | 17.1 | 5 | 8 | 27.6 | 3.1 | 53 | 29 | 20.5 |
| | M | 767 | 74.8 | 439 | 74 | 57.2 | 63 | 82.9 | 8.2 | 21 | 72.4 | 2.7 | 130 | 71 | 16.9 |

(*Continued*)

**Table 3.** *Continued.*

| | | Scientists: total | | Scientists: general collaboration | | | Scientists: international collaboration | | | Scientists: national collaboration | | | Scientists: institutional collaboration | | |
|---|---|---|---|---|---|---|---|---|---|---|---|---|---|---|---|
| | | *N* | % col | *N* | % col | % row | *N* | % col | % row | *N* | % col | % row | *N* | % col | % row |
| PHARM | F | 169 | 66.5 | 166 | 66.4 | 98.2 | 3 | 50 | 1.8 | 9 | 60 | 5.3 | 129 | 66.2 | 76.3 |
| | M | 85 | 33.5 | 84 | 33.6 | 98.8 | 3 | 50 | 3.5 | 6 | 40 | 7.1 | 66 | 33.8 | 77.6 |
| PHYS | F | 182 | 16.6 | 162 | 16.4 | 89 | 25 | 14.5 | 13.7 | 10 | 26.3 | 5.5 | 85 | 17 | 46.7 |
| | M | 916 | 83.4 | 824 | 83.6 | 90 | 147 | 85.5 | 16 | 28 | 73.7 | 3.1 | 414 | 83 | 45.2 |
| BUS | F | 372 | 52.1 | 237 | 51.7 | 63.7 | 43 | 53.8 | 11.6 | 23 | 53.5 | 6.2 | 125 | 50.8 | 33.6 |
| | M | 342 | 47.9 | 221 | 48.3 | 64.6 | 37 | 46.3 | 10.8 | 20 | 46.5 | 5.8 | 121 | 49.2 | 35.4 |
| DENT | F | 57 | 76 | 57 | 76 | 100 | 0 | 0 | 0 | 1 | 50 | 1.8 | 42 | 76.4 | 73.7 |
| | M | 18 | 24 | 18 | 24 | 100 | 0 | 0 | 0 | 1 | 50 | 5.6 | 13 | 23.6 | 72.2 |
| ECON | F | 186 | 49.1 | 92 | 46.2 | 49.5 | 10 | 62.5 | 5.4 | 12 | 44.4 | 6.5 | 42 | 47.2 | 22.6 |
| | M | 193 | 50.9 | 107 | 53.8 | 55.4 | 6 | 37.5 | 3.1 | 15 | 55.6 | 7.8 | 47 | 52.8 | 24.4 |
| HEALTH | F | 23 | 34.3 | 19 | 33.3 | 82.6 | 0 | 0 | 0 | 4 | 80 | 17.4 | 7 | 24.1 | 30.4 |
| | M | 44 | 65.7 | 38 | 66.7 | 86.4 | 0 | 0 | 0 | 1 | 20 | 2.3 | 22 | 75.9 | 50 |
| HUM | F | 527 | 49.8 | 99 | 50 | 18.8 | 17 | 43.6 | 3.2 | 16 | 50 | 3 | 52 | 48.1 | 9.9 |
| | M | 531 | 50.2 | 99 | 50 | 18.6 | 22 | 56.4 | 4.1 | 16 | 50 | 3 | 56 | 51.9 | 10.5 |
| MED | F | 1920 | 53.7 | 1862 | 53.5 | 97 | 44 | 46.3 | 2.3 | 89 | 54.9 | 4.6 | 1351 | 54.4 | 70.4 |
| | M | 1654 | 46.3 | 1621 | 46.5 | 98 | 51 | 53.7 | 3.1 | 73 | 45.1 | 4.4 | 1132 | 45.6 | 68.4 |
| PSYCH | F | 194 | 63.8 | 148 | 63 | 76.3 | 16 | 57.1 | 8.2 | 16 | 76.2 | 8.2 | 50 | 60.2 | 25.8 |
| | M | 110 | 36.2 | 87 | 37 | 79.1 | 12 | 42.9 | 10.9 | 5 | 23.8 | 4.5 | 33 | 39.8 | 30 |
| SOC | F | 494 | 49.8 | 193 | 51.5 | 39.1 | 23 | 41.1 | 4.7 | 17 | 43.6 | 3.4 | 120 | 56.6 | 24.3 |
| | M | 498 | 50.2 | 182 | 48.5 | 36.5 | 33 | 58.9 | 6.6 | 22 | 56.4 | 4.4 | 92 | 43.4 | 18.5 |
| VET | F | 146 | 44 | 143 | 43.6 | 97.9 | 0 | 0 | 0 | 4 | 44.4 | 2.7 | 122 | 44.7 | 83.6 |
| | M | 186 | 56 | 185 | 56.4 | 99.5 | 1 | 100 | 0.5 | 5 | 55.6 | 2.7 | 151 | 55.3 | 81.2 |
| Total | F | 10,577 | 41.5 | 8803 | 42.3 | 83.2 | 438 | 36.2 | 4.1 | 478 | 45.3 | 4.5 | 5668 | 44.6 | 53.6 |
| | M | 14,886 | 58.5 | 12,003 | 57.7 | 80.6 | 772 | 63.8 | 5.2 | 577 | 54.7 | 3.9 | 7033 | 55.4 | 47.2 |

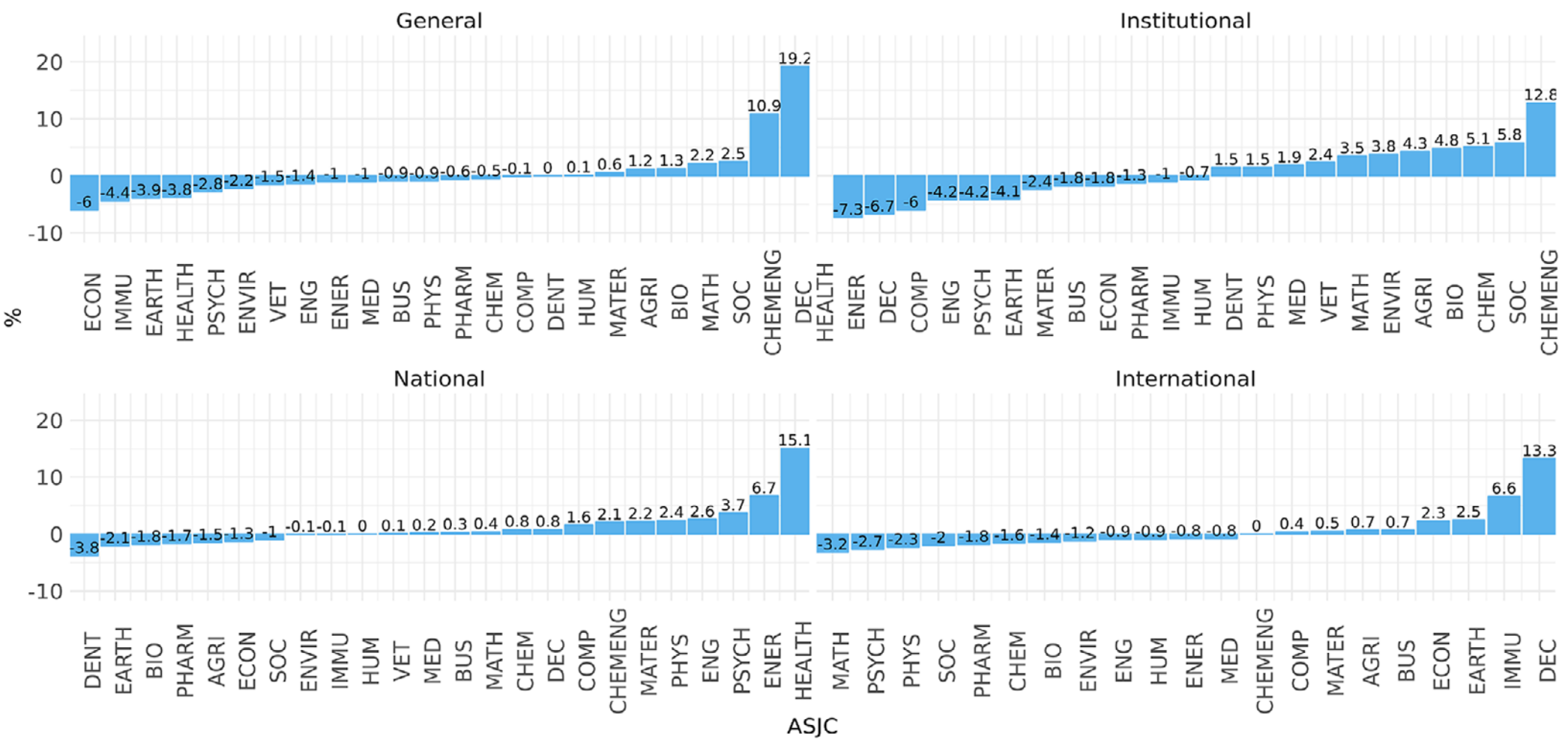

**Figure 4.** Percentage Point Gender Differences in High-Intensity Collaboration (>75% Articles Published in the Four Collaboration Types in the Scientist's Individual Publication Portfolio for 2009–2018) for All Polish Internationally Visible University Professors, by Collaboration Type (Four Panels), and ASJC Discipline. *Note*: Vertical axes: positive results indicate female advantage in a given ASJC discipline, negative results indicate male advantage, and 0 indicates exactly the same distribution of collaboration by gender. [Colour figure can be viewed at wileyonlinelibrary.com]

disciplines of IMMU and DEC with female advantage. It is only in selected disciplines that over 10% of scientists reach this high-intensity international collaboration level: interestingly, it is BUS in the general cluster of non-STEM fields (both females and males, 11.6% and 10.8%) and PHYS in the general cluster of STEM fields (both females and males, 13.7% and 16.0%). In addition, this collaboration intensity was found for males in PSYCH and females in DEC.

To summarize, gender disparities in collaboration have to be studied by collaboration type. They are different for the four collaboration types analyzed. While the propensity to engage in general, national, and institutional collaboration is higher for female Polish scientists, the propensity to collaborate internationally is higher for male Polish scientists. However, as analyzed in detail above, the differences are not substantial and there are significant cross-disciplinary gender differentiations. Notably, for each collaboration type, there are specific disciplines in which the above overall picture does not fit the picture disaggregated to the level of disciplines.

Finally, gender disparities in high-intensity international collaboration at the disaggregated level of disciplines can also be examined from another perspective: disciplines with female advantage would be those in which the share of female scientists involved in high-intensity international collaborations would be higher than the overall share of female scientists in these disciplines. In the case of no gender disparities, the percentages of all females and females with high intensity in international collaboration must be exactly the same. Figure 5 indicates that there are eight such disciplines, including one of the three largest (AGRI) and the middle-sized STEM disciplines of COMP, EARTH, and MATER. Apart from CHEMENG (with no gender disparity viewed from this angle), an advantage for males was found for all other disciplines, including in large disciplines such as MED and ENG, non-STEM disciplines such as HUM and SOC, and STEM disciplines such as PHYS, MATH, BIO, and CHEM. In this section, a small-sized discipline ($n = 481$ scientists) of chemical engineering (CHEMENG), as studied from several angles, clearly emerges as a prototypical discipline with no gender disparities in international research collaboration. ECON (Economics, Econometrics, and Finance), of specific interest to readers of this journal, is summarized separately in Section 4.4.

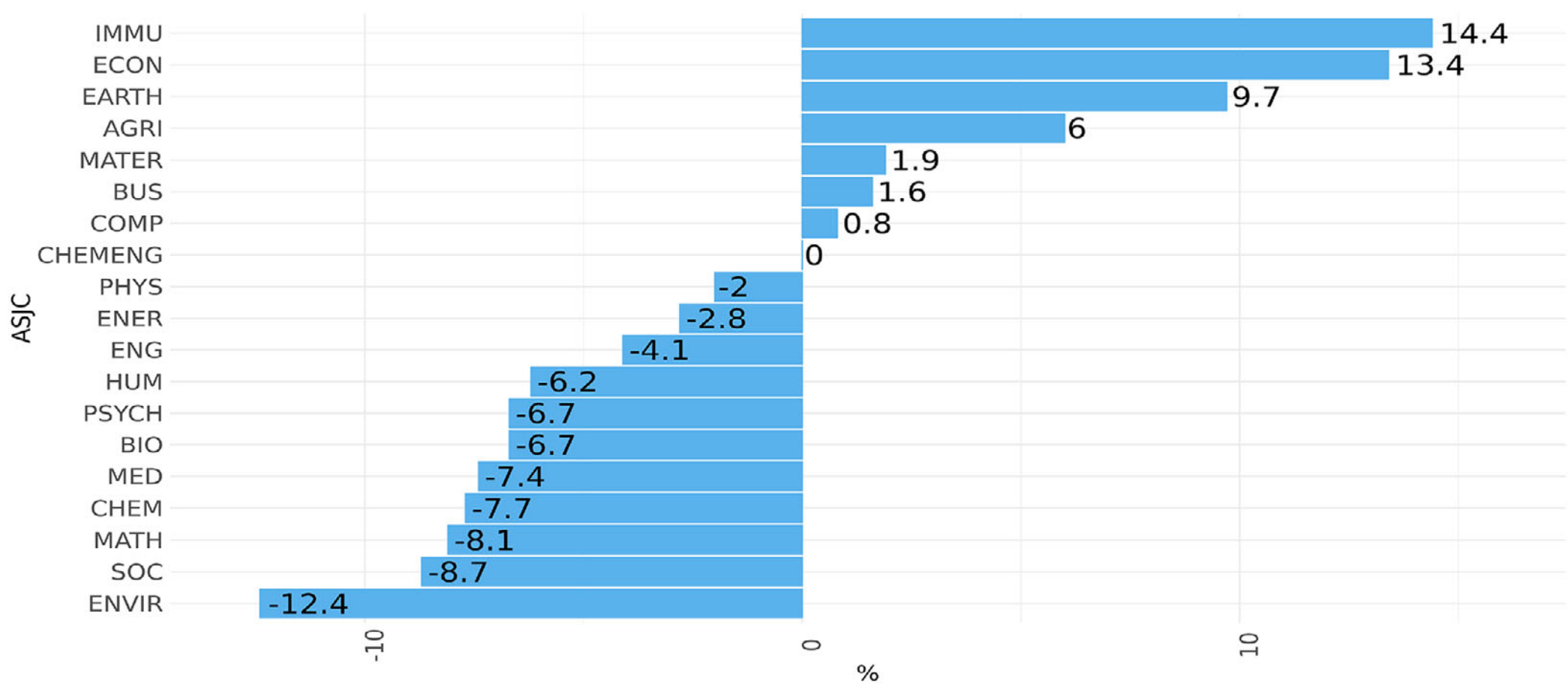

**Figure 5.** Percentage Point Differences Between the Overall Share of Female Scientists and the Share of Female Scientists Involved in High-Intensity International Collaboration, All Polish Internationally Visible University Professors, by ASJC Discipline (Five Disciplines Omitted: Low Counts) (in %). [Colour figure can be viewed at wileyonlinelibrary.com]

### 4.3 *Gender Distribution in International Collaboration: Cross-Disciplinary Differences*

In the next stage of analysis, it is useful to present cross-disciplinary distribution differences in international collaboration by gender using boxplots. The first one is for all internationally visible Polish university professors (both collaborating and noncollaborating internationally—a total of 25,463—Figure 6); and the second is only for a subsample of university professors collaborating internationally (that is, with at least a single article written in international collaboration—11,854 or 46.6%—within the decade studied, Figure 7). Thus, the international collaboration examined here is one that is beyond the three selected intensity levels. Scientists in disciplines such as PHYS, followed by BIO and CHEM, for both males and females, represent the highest average level of international collaboration (in their individual publication portfolios, Figure 6). For example, for female scientists in PHYS, 50% of authors have a share of at least 25% in internationally coauthored articles and 50% of authors have at most a 25% share of this publication type in their individual portfolios. In all three cases, the median value for males is higher than the median value for females. For HUM, ECON, SOC, and DENT in the cluster of non-STEM fields, all quantiles up to the third quartile equal zero for both males and females. This effectively implies, as evident from Figure 6, that all observations in these disciplines are outliers—that is, there are a small number of scientists with atypically high shares of internationally coauthored articles in their individual publication portfolios within these disciplines.

The average level of intensity of international collaboration (at the level of individual scientists) by gender in the case of a subsample of internationally collaborating scientists is presented in Figure 7. The median level reached approximately 50% for males in such disciplines in the cluster of non-STEM fields as HUM, BUS, PSYCH, and SOC and in disciplines in the cluster of STEM fields such as COMP, PHYS, and MATH. For females, the median reached similar levels in HUM, BUS, PSYCH, and SOC, as well as in COMP, but not in PHYS and MATH. However, it is important to note relatively low numbers of scientists collaborating internationally in some disciplines (from 1,875 in MED and 1,362 in AGRI to 14 in DENT). To summarize, as a general pattern, the average shares of internationally collaborative papers in the individual publication portfolios of male scientists are higher or equal to those of female scientists although in most cases differences are not striking.

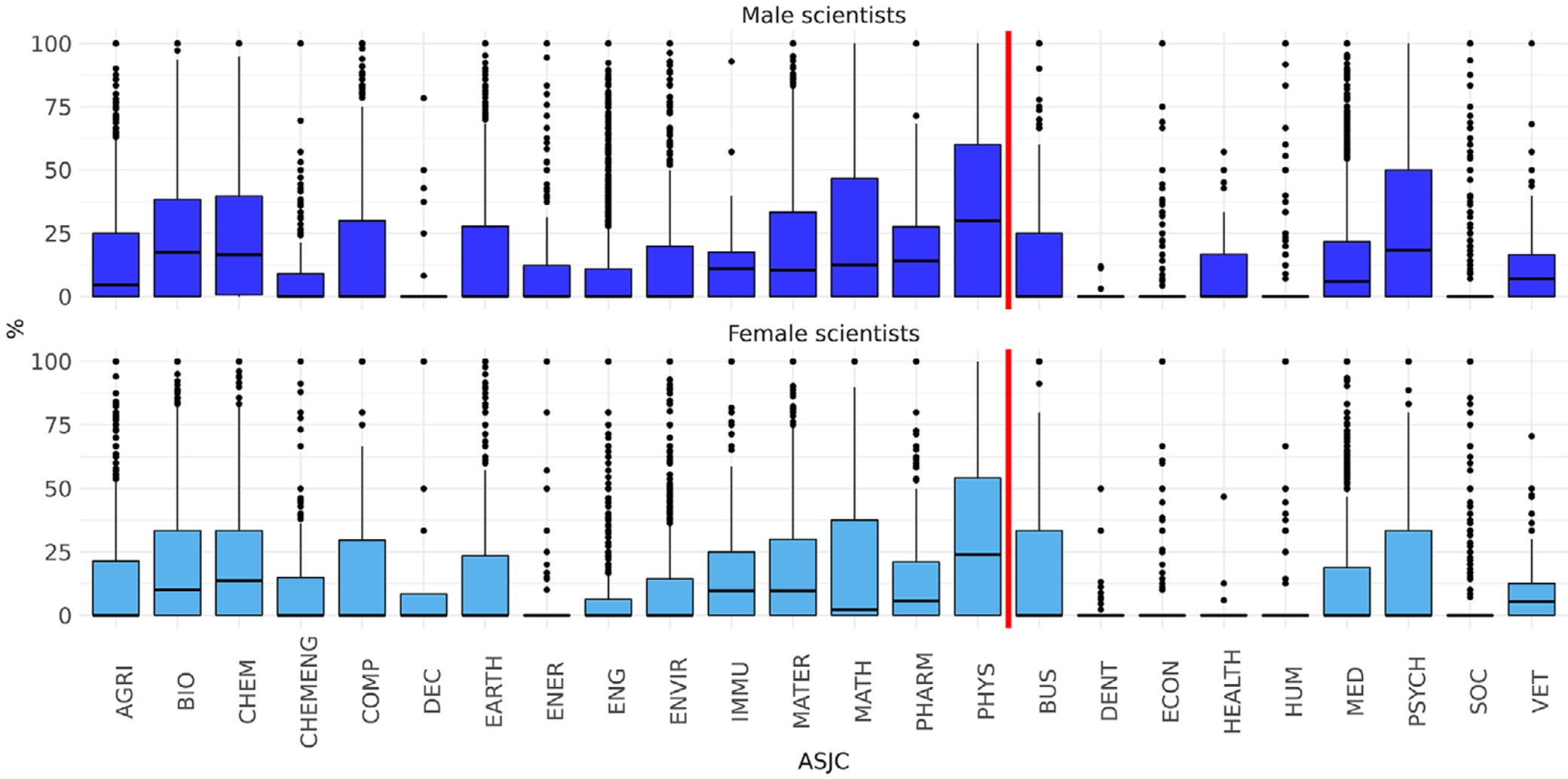

**Figure 6.** Distribution of International Collaboration Percentages in Scientists' Individual Publication Portfolios for 2009–2018, All Polish Internationally Visible University Professors ($N = 25{,}463$, Both Collaborating and Noncollaborating Internationally), by Gender and ASJC Discipline (in %). STEM (left panels) and Non-STEM (right panels) ASJC Disciplines. [Colour figure can be viewed at wileyonlinelibrary.com]

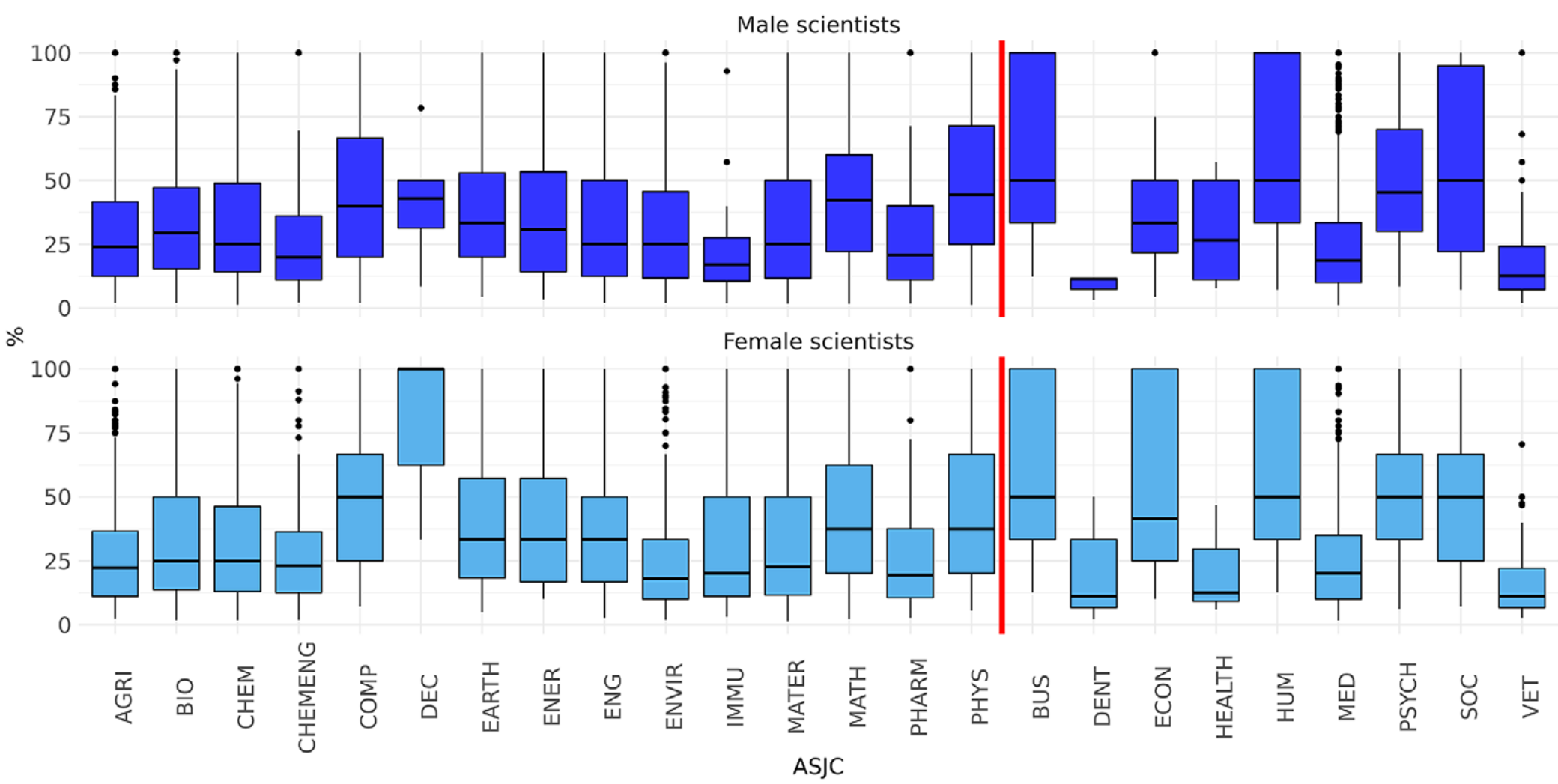

**Figure 7.** Distribution of International Collaboration Percentages In Scientists' Individual Publication Portfolios for the Period 2009–2018, Internationally Collaborating University Professors Only ($N = 11{,}854$; Low Collaboration Intensity), by Gender and ASJC Discipline (in %). STEM (Left Panel) and Non-STEM (Right Panel) ASJC Disciplines. [Colour figure can be viewed at wileyonlinelibrary.com]

**Table 4.** Distribution of International Collaboration for the Period 2009–2018, Internationally Collaborating University Professors Only ($N = 11{,}854$), by Gender, Age Group, and Collaboration Intensity. All ASJC Disciplines Combined.

| | | Young (39 and younger) | | Middle aged (40–54) | | Older (55 and older) | | Total | |
|---|---|---|---|---|---|---|---|---|---|
| | | *n* | % | *n* | % | *n* | % | *n* | % |
| Female scientists | Low | 1636 | 45.72 | 2286 | 44.53 | 875 | 46.92 | 1636 | 45.72 |
| | Middle | 344 | 9.61 | 395 | 7.69 | 149 | 7.99 | 344 | 9.61 |
| | High | 178 | 4.97 | 184 | 3.58 | 76 | 4.08 | 178 | 4.97 |
| Male scientists | Low | 1811 | 48.33 | 2995 | 45.89 | 2251 | 48.80 | 1811 | 48.33 |
| | Middle | 456 | 12.17 | 619 | 9.49 | 527 | 11.42 | 456 | 12.17 |
| | High | 218 | 5.82 | 292 | 4.47 | 262 | 5.68 | 218 | 5.82 |

### 4.4 *Gender, Collaboration, and Age Distribution*

While we generally refer to individual publication portfolios constructed for every Polish male and female scientist (and therefore do not refer to the time dimension), in this section (and in Section 4.1), we refer to the age distribution in international research collaboration. The time dimension could not be captured through portfolios as the number of publications per scientist per year was too small for a meaningful gender comparison. We therefore studied aggregates: individual outputs of 10 years (or less in the case of the newly employed).

We will examine here the changing relationship between age (as of 2017, a point in time) and the share of internationally collaborating scientists of that age. Table 4 shows how this contingency is reflected in the international collaboration rate by gender, age group, and collaboration intensity. For instance, the share of young female scientists involved in low-intensity collaboration in the total population of young female scientists is 45.72%, compared with 48.33% for young male scientists. Table 4 allows us to compare the involvement of the three age groups of scientists at the three collaboration intensity levels by gender: for instance, the share of female scientists involved in international collaboration at a high intensity level across age groups, from young to middle-aged to older scientist (which ranges from 3.58% to 4.97%). The visualization of this joint distribution (for the three age gropus and all 24 ASJC disciplines combined) in Figure 8 highlights that the differences by age group are similar for male and female scientists: while for low collaboration intensity, the collaboration rate is the lowest for scientists in the second age group (40–54), for both middle and high intensities, it is the highest for young scientists. Gender disparities are small for low collaboration intensity but much more significant for middle and high intensities (compare the left and the right panels in Figure 8).

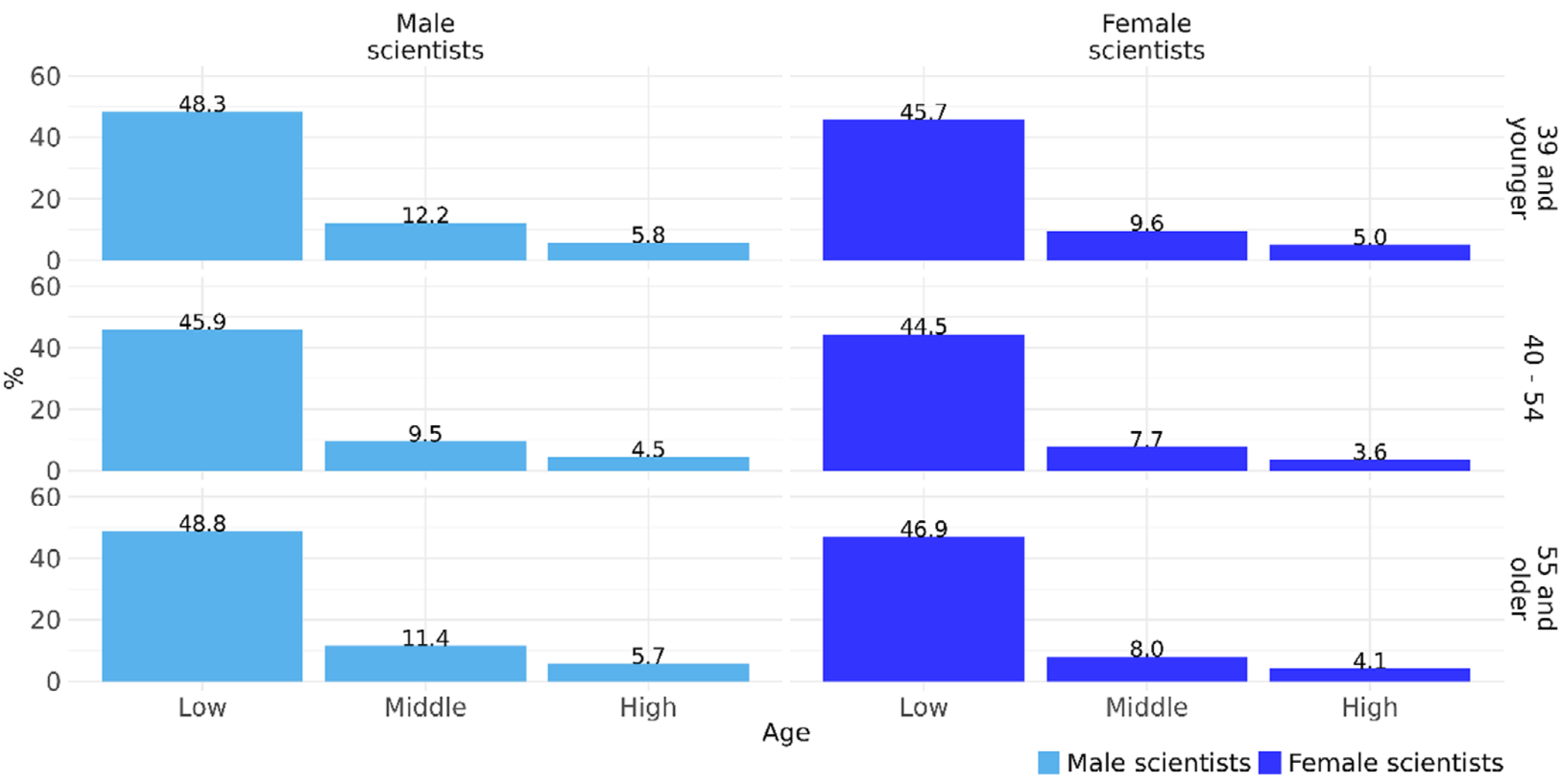

**Figure 8.** Distribution of international collaboration percentages for the period 2009–2018, internationally collaborating university professors only (N = 11,854), by gender, age group, and collaboration intensity. All ASJC disciplines combined. [Colour figure can be viewed at wileyonlinelibrary.com]

However, the relationship can be examined in more detail: first, by using the joint distribution of all ASJC disciplines combined, in which every year of age is shown (Figure 9) so that male and female scientists are examined by yearly age groups, and, second, by using the distribution in which each ASJC discipline is examined separately (Figure 10).

The regression lines in Figure 9 illustrate relationships between age and international collaboration rate. For each age, the rate is higher for males than for females (exceptions include university professors older than 65, with a relatively low number of female observations). However, the gray areas representing 95% confidence intervals overlap for some ages (visualized as dark gray areas): in the majority of cases, the gender differences are not statistically significant, especially for low- and high-intensity collaboration. In Figure 10, the relationships are simplified: a linear relationship between age and collaboration rate is assumed for the purposes of our analysis, and cross-disciplinary differences are shown. The changes in the share of internationally collaborating scientists are shown by age and contrasted by gender for STEM (top panel) and non-STEM (bottom panel) disciplines. There are clearly disciplines in which males of all ages are consistently more internationalized (for instance, BIO, CHEM, or HEALTH). While females of all ages are consistently more internationalized in only one case (CHEMENG), in several cases, older females (aged 50 and older) are more internationalized than older males (MATH, PHYS, and SOC). The emergent picture is highly differentiated: in some disciplines, the internationalization rate increases with age either for both genders or for one gender only; in other disciplines, similarly, the rate decreases with age for both genders or one gender only. Thus, all combinations of gender-related increase and decrease are represented.

## 4.5 *Economics, Econometrics, and Finance*

The patterns were also tested for ECON (Economics, Econometrics, and Finance), with an almost equal gender distribution (49.1% of our sample of 379 economists were female). The percentage point

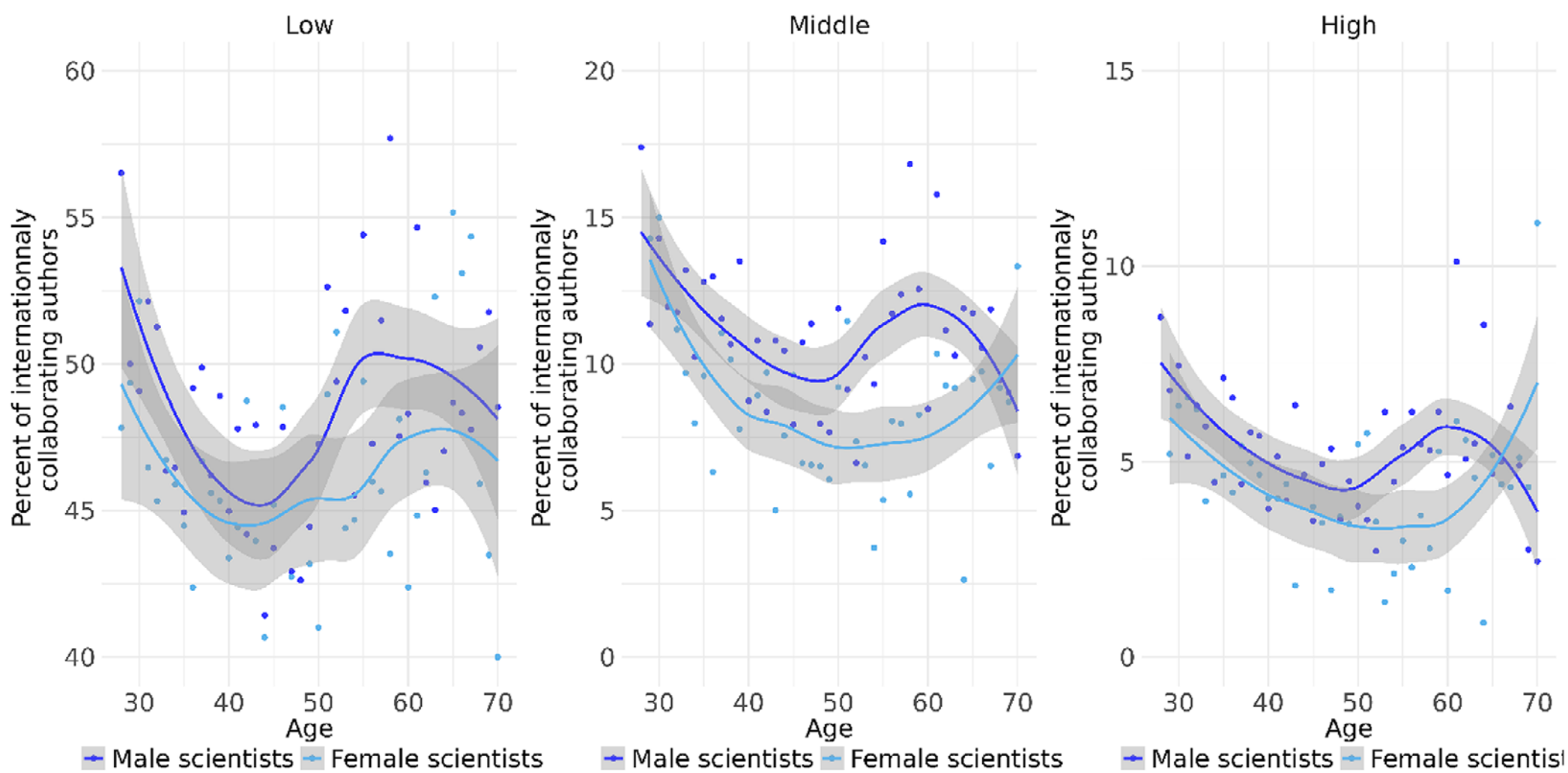

**Figure 9.** The Percentage of Internationally Collaborating University Professors by Gender, Age, and Collaboration Intensity ($N = 25{,}463$). All ASJC Disciplines Combined. The Regression Line Was Estimated Using the Method of Local Polynomial Regression Fitting. Gray Areas Represents 95% Confidence Intervals. Each Year of Age Is Represented by a Single Dot (a Cutoff Point of the Age of 70 Used). Dots Represent the Percentage of University Professors of a Particular Age. [Colour figure can be viewed at wileyonlinelibrary.com]

gender difference in high-intensity collaboration for economists indicates a male advantage in the case of general, institutional, and national collaboration but—interestingly—a female advantage in the case of international collaboration. The male advantage in the case of general collaboration is the highest of all disciplines studied (6 p.p.; see Figure 4). The picture for economists in the case of high-intensity international collaboration is clear: female economists are more internationally collaborative in their individual publication portfolios than males, against the opposite aggregated picture of all disciplines combined, in which males are more internationally collaborative. At the same time, the percentage of internationally collaborating economists is higher for males than for females of all ages and does not exceed 25% (Figure 10), a picture that is similar to arts and humanities (HUM). Economics is also one of only seven disciplines for which the share of females involved in high-intensity international research collaboration is higher than the overall share of females (by 13.4 p.p.; see Figure 5), following only the scantly represented discipline of IMMU. Finally, referring to the model below (Table 5), for ECON, the likelihood of being internationally collaborative is twice as high as for arts and humanities taken as a reference category.

### *4.5.1 A Model Approach: Logistic Regression Analysis*

In our modeling approach, the strength of the joint effect of traditional predictors of international research collaboration was tested. In this section, international research collaboration is understood as a low-intensity collaboration: having at least one article published in international collaboration in one's individual publication portfolio in the study period of 10 years from 2009 to 2018.

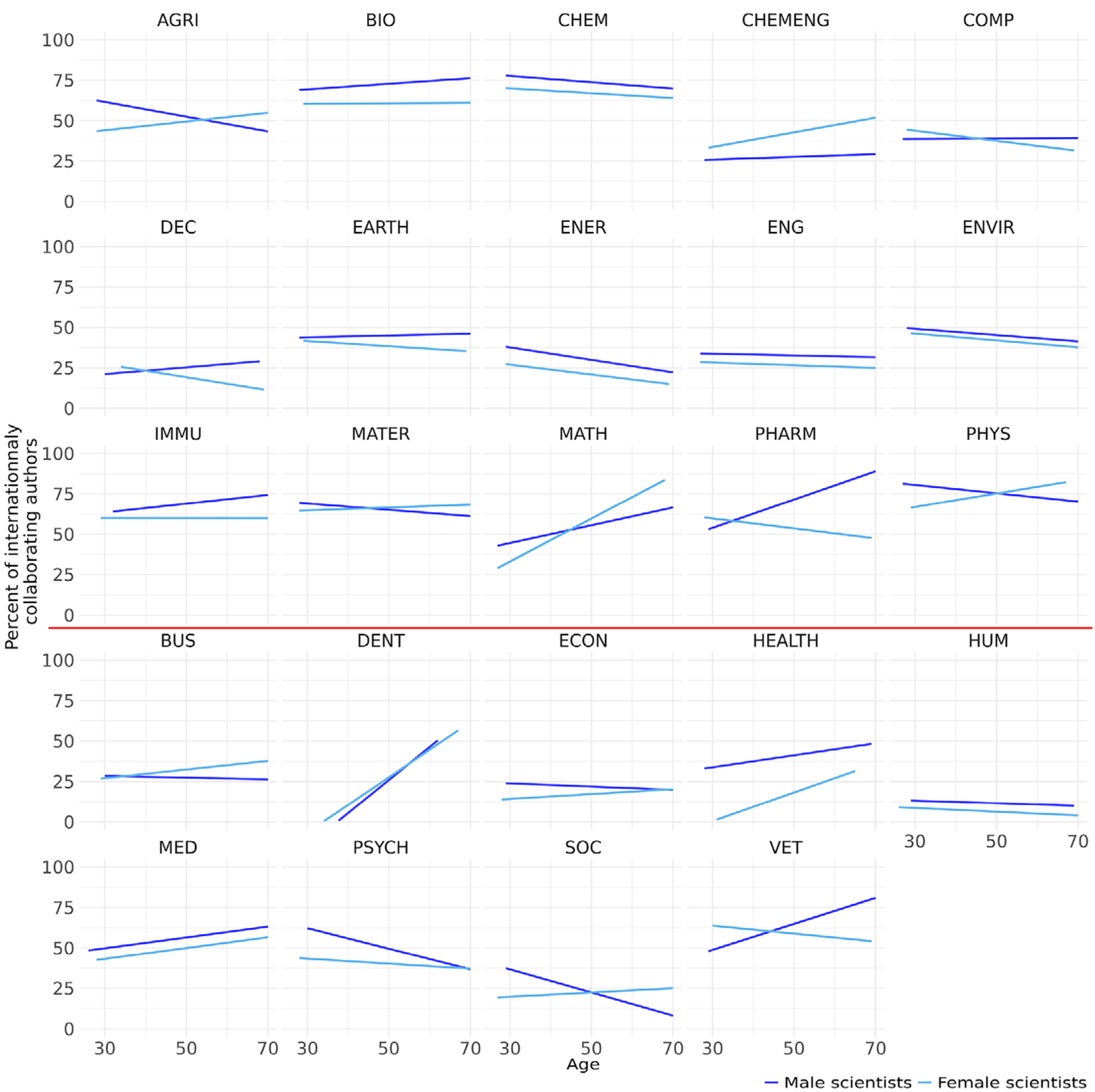

**Figure 10.** The Percentage of Internationally Collaborating University Professors, by Gender, Age, and ASJC Discipline. Low-Intensity Collaboration Only ($N = 11{,}854$). For Age, a Cutoff Point of 70 Is Used. The Red Line Separates STEM from Non-STEM Disciplines. [Colour figure can be viewed at wileyonlinelibrary.com]

An analytical linear logistic model was constructed based on research literature, particularly predictive models built in Cummings and Finkelstein (2012), Rostan *et al.* (2014), Sooryamoorthy (2014), and Finkelstein and Sethi (2014). Three models were built: Model 1 for all scientists, Model 2 for male scientists, and Model 3 for female scientists. Estimating the odds ratios of being scientists defined as “internationally collaborative” was based on a set of independent variables: age, gender (reference category: female), academic position (reference category: assistant professor or the lowest position in our study), institutional type (reference category: the university or the institutional type with the lowest share of internationally collaborative articles, with traditional comprehensive universities having

**Table 5.** Odds Ratio Estimates of Being Internationally Collaborative, Three Logistic Regression Models.

| | Model 1 All scientists $R^2 = 0.355$ | Model 2 Male scientists $R^2 = 0.381$ | Model 3 Female scientists $R^2 = 0.320$ |
|---|---|---|---|
| Age | 0.983*** | 0.984*** | 0.982*** |
| Male | 1.124*** | | |
| Associate professor (degree: Dr. hab.) | 1.183*** | 1.206*** | 1.152* |
| Full professor (title: professor) | 1.511*** | 1.55*** | 1.352** |
| Classical (comprehensive) university | 1.496*** | 1.485*** | 1.504*** |
| Technical university | 0.962 | 0.876* | 1.122 |
| Total productivity - number of articles | 1.123*** | 1.128*** | 1.118*** |
| Field: AGRI | 4.552*** | 3.977*** | 5.304*** |
| Field: BIO | 7.588*** | 8.218*** | 7.763*** |
| Field: CHEM | 7.419*** | 6.998*** | 7.963*** |
| Field: CHEMENG | 3.145*** | 2.503*** | 4.146*** |
| Field: COMP | 4.764*** | 4.441*** | 5.866*** |
| Field: DEC | 3.322*** | 2.504 | 4.584** |
| Field: EARTH | 4.226*** | 3.928*** | 4.537*** |
| Field: ENER | 3.23*** | 3.286*** | 2.763** |
| Field: ENG | 2.873*** | 2.762*** | 2.893*** |
| Field: ENVIR | 3.257*** | 3.098*** | 3.447*** |
| Field: IMMU | 6.946*** | 6.021*** | 8.051*** |
| Field: MATER | 6.086*** | 5.475*** | 6.948*** |
| Field: MATH | 6.357*** | 5.779*** | 7.077*** |
| Field: PHARM | 4.653*** | 5.565*** | 4.752*** |
| Field: PHYS | 10.041*** | 8.899*** | 13.027*** |
| Field: BUS | 5.029*** | 3.889*** | 6.333*** |
| Field: DENT | 1.039 | 0.461 | 1.471 |
| Field: ECON | 2.174*** | 1.956** | 2.468*** |
| Field: HEALTH | 2.97*** | 3.961*** | 1.121 |
| Field: MED | 3.989*** | 3.505*** | 4.645*** |
| Field: PSYCH | 5.146*** | 6.068*** | 5*** |
| Field: SOC | 2.334*** | 2.104*** | 2.652*** |
| Field: VET | 4.999*** | 4.634*** | 5.46*** |
| Constant | 0.111*** | 0.128*** | 0.106*** |

*$p < 0.05$; **$p < 0.01$; ***$p < 0.001$

the highest share of them), productivity (the total number of articles indexed in the Scopus database in the study period 2009–2018), academic disciplines (defined as dominant Scopus ASJC categories, reference category: HUM, that is, arts and humanities, or the reference category with the lowest share of internationally collaborative articles in Poland). Importantly, all data in the models come from the integrated database—that is, originally from an official administrative and biographical database and a Scopus journal publication and citation database; consequently, the data are as objective as they can be (no self-declared data as in academic profession surveys are used in this model, thereby making its results more robust). The total number of observations used in the model was 25,463 (10,577 or 41,5% females and 14,886 or 58.5% males). The occurrence of potential multicollinearity was tested using an inverse correlation matrix. From among the variables analyzed, age turned out to be significantly correlated with a vector of other independent variables; however, age was entered into the model with the awareness that the estimate efficiency of this parameter was reduced.

For all scientists (Model 1), the model fit rather well to the data, as Nagelkerke's $R^2 = 0.36$. The data indicated that being a male scientist increases the chances of being internationally collaborative by merely 12.4% on average compared with female scientists (all other parameters being equal, see the details in Table 5). Interestingly, each year of age decreases that chance by 1.7% on average. In general, this is in line with descriptive statistics. The implication is that younger scientists are more prone to collaborating internationally. Being an associate professor increases the odds by approximately one-fifth compared with being assistant professor, and being a full professor increases that chance by half.

Further, working in a classical (comprehensive) university increases the odds by half compared with working in a university (defined as a noncomprehensive, such as university of economics or medical university). There were no differences in the odds for technical universities in the same comparison (comprehensive vs. noncomprehensive). The higher the total individual productivity, the greater the odds of being internationally collaborative. Each (Scopus-indexed) article published increases the odds by as much as 12.3%. Almost all disciplines (except dentistry) are characterized by significantly larger odds compared with arts and humanities. The most internationally collaborative disciplines are physics and astronomy (PHYS, with scientists 10 times more likely to collaborate than academics in arts and humanities), biochemistry, genetics and molecular biology, and chemistry (BIO and CHEM, both over seven times more likely). The least internationally collaborative disciplines are engineering (ENG) and health professions (with scientists three times more likely to collaborate than academics in arts and humanities).

Global literature on gender disparities in international research collaboration indicates that separate regression models for each gender might be useful. Certain variables might have a much stronger effect on only male or only female scientists, while others do not differentiate by gender. In our case, apparently, being a male full professor increases the odds of being internationally collaborative by half, while for females it increases the odds by approximately one-third. Further, publishing in the field of business, management, and accounting (BUS) increases the odds three times for males compared with five times for females (Models 2 and 3). However, prior to any further analysis, a hypothesis of a statistically significant difference between parameter values in the two models was tested. This hypothesis was tested by comparing 95% confidence intervals of logistic regression parameters for Model 2 (male scientists) and Model 3 (female scientists). If the intervals overlapped to any extent, it implied that the difference between parameter values is not significantly different from zero. This effectively implies that the parameter values in the population studied are equal to each other and no gender difference can be shown. Confidence intervals for each parameter in regression models are indeed overlapping. In a very specific Polish case, a modeling approach to examining gender disparities in international research collaboration must result in a reliance on a single holistic model (Model 1) with gender as an independent variable.

Our unit of analysis is an individual scientist with their clearly defined individual publication portfolio for the whole period studied; consequently, the time dimension per scientist cannot be examined due to low publication counts per year. Our study compares the aggregates of publication portfolios by male and female scientists. At the national level, however, Scopus data clearly indicate that the percentage of internationally collaborative papers in national research output was steadily increasing in the study period of 2009–2018, from 29.1% to 35.8 %.

## 5. Summary of Findings, Discussion, and Conclusions

This research reveals substantial gender disparities in the collaboration patterns of Polish scientists. This represents the first time that such differences have been systematically explored from a large-scale bibliometric perspective (using the Polish Science Observatory database maintained by the authors). A detailed examination of our administrative, biographical, publication, and citation database of all internationally visible Polish university professors ($N = 25{,}463$, including 158,743 journal articles written in the decade 2009–2018 by 14,886 male and 10,577 female scientists) leads us to a number of conclusions.

First, while female scientists exhibit a higher rate of general, national, and institutional collaboration, male scientists exhibit a higher rate of international collaboration. Gender differences are not high; however, they are statistically significant for all four major research collaboration types. This finding is critically important in explaining gender disparities in terms of impact, productivity, and access to large grants, in view of the fundamental role of international collaboration in global science in comparison with any other collaboration type (Larivière *et al*., 2011; Gazni *et al*., 2012; Wagner, 2018). Initially small gender differences in international research collaboration accumulate over time, contributing to the widening of gender impact, productivity, and research funding gaps. These findings are in line with literature that emphasizes the stable nature of gender gaps in science despite increasing female participation in it.

However, second, an aggregated picture of gender disparities in international research collaboration hides a much more nuanced picture of gender disparities by discipline—and this aspect of our findings calls for special attention. There are substantial cross-disciplinary gender differentiations in international research collaboration (and in the three other collaboration types examined, as Thelwall and Maflahi, 2019, show for national collaboration); furthermore, there are specific disciplines in which the above overall picture of male advantage in international collaboration does not fit the picture when disaggregated to the level of certain disciplines, with a clear female advantage in these disciplines. Notably, these disciplines include computer science (COMP); business, management, and accounting (BUS); economics, econometrics, and finance (ECON); agricultural and biological sciences (BIO); and earth and planetary sciences (EARTH).

Therefore, we can conclude that (1) male scientists exhibit higher collaboration rates in only one collaboration type—international collaboration; in all other collaboration types (general, national, and institutional), female scientists are more collaborative; (2) there is no one-size-fits-all answer to the question of gender disparity in international collaboration: differences by discipline are fundamental but remain hidden in aggregated data; and (3) the power of individual-level data (with the scientist as the unit of analysis) is underestimated, and data sets for entire populations of scientists in other national systems are required to further explore collaboration patterns. These findings are in line with the general conclusions drawn in Abramo *et al*. (2013), which focused on Italian scientists, and with the global conclusions reached by Larivière *et al*. (2013). In contrast, they are not in line with the findings in Aksnes *et al*. (2019) regarding international collaboration. Aksnes and colleagues have also shown the power of the field of research in influencing the propensity to collaborate internationally, but in the Norwegian case, when field, academic rank, and productivity were taken into consideration, there were only minor and non–statistically significant sex differences in the propensity to collaborate internationally. In sum, both discipline-focused global research and other national-level studies are needed to explore the topic further.

Third, we examined international research collaboration at three separate intensity levels (low, medium, and high), with male scientists dominating in international collaboration at each. This male domination is systemic but not substantial (for instance, while 47.4% of male scientists collaborate internationally at the low intensity level, 10.8% at the medium intensity level, and 5.2% at the high intensity level, for female scientists, the rates are 45.4%, 8.4%, and 4.1%, respectively). However, interestingly, at each intensity level, there are specific disciplines in which females collaborate more than males: for example, at the high-intensity collaboration level, female scientists have higher collaboration rates in nine disciplines. There are also eight disciplines, including one of the three largest (AGRI, representing more than one in ten Polish scientists), in which females engaging in high-intensity international collaboration are overrepresented compared with males. Moreover, chemical engineering, which is a discipline represented by a relatively small group of scientists ($n = 481$), emerges as a discipline with no gender disparities in international research collaboration from whatever angle it is examined. All these disciplines together are populated by approximately 8,000 internationally visible university professors, or by almost one-third of all those in our sample.

Fourth, having the date of birth of every scientist, we examined the relationship between age and the share of internationally collaborating scientists. This is where we went far beyond previous global and national studies.

Three age groups and then all ages were compared for females and males, both for all disciplines combined and for each discipline separately. Gender disparities are small for low collaboration intensity and much more significant for middle and high intensity. For each age, the share of males collaborating internationally among all males of that age is higher than the same rate for females—but overlapping 95% confidence intervals indicate that the differences are not statistically significant. For all three collaboration intensities, the share of collaborating scientists is the highest for very young scientists around 30 years of age and the lowest for those around 40–50. The emergent picture is highly differentiated, and all combinations of gender-related increase and decrease are represented. The share of male and female scientists involved in international collaboration generally decreases with age until the early 40s; then, it increases abruptly for males until about 55 and gently for females until about 65. Our major cross-gender finding is that there are two peaks in the population studied in the share of international research collaboration: scientists around 30 and 60 years of age—and males are more highly involved in international collaboration at statistically significant levels only in the latter case. The Polish system is thus characterized by small gender differences in the share of internationally collaborative scientists until about the age of 50; thereafter, these grow, substantially increasing for scientists in their 50s and 60s, when higher proportions of internationally collaborative males are accompanied by lower proportions of internationally collaborative females, with high cross-disciplinary differentiation.

The following conclusion can be drawn from linear logistic models: in the Polish case, a single holistic model with gender as an independent variable works better than separate models for the two genders. Somehow, surprisingly, being a male scientist increases the odds of being internationally collaborative by merely 12.4% (the odds would increase if being "highly collaborative" were the dependent variable). Further, age, academic position, institutional type, total productivity, and working in selected disciplines are significant. Age decreases the odds, as expected (each year of age by 1.7% on average). The likelihood of being internationally collaborative increases by half for full professors and by approximately one-fifth for associate professors, as it does for scientists working in classical (comprehensive) universities. The higher the total individual productivity, the greater the odds of being internationally collaborative (each article published increasing the odds by as much as 12.3%). Furthermore, compared with arts and humanities, the likelihood also abruptly increases for scientists from such STEM fields as physics and astronomy (as much as 10 times), as well as from biochemistry, genetics, molecular biology, and chemistry (seven times).

The current study is comprehensive and examines international collaboration in the context of all other collaboration types. By using a data set with a combination of administrative, biographical, and bibliometric data for all internationally visible Polish scientists, the study goes beyond bibliometrics. The methodological stance of using the individual scientist (each with an individual publication portfolio) as the unit of analysis, the study avoids the pitfalls of aggregation and overreliance on highly productive, highly internationalized scientists present in every system (and in Poland, 10% of scientists produce 50% of all publications, Kwiek, 2018b). As the unit of analysis in our study is a single scientist, the role of a female scientist with five internationally collaborative articles (out of 10) in detecting collaboration patterns is exactly the same as the role of the one with 50 publications (out of 100): for both observations, the international collaboration intensity is 50%.

Policy implications of this research are straightforward: first, given that gender disparities in international research collaboration vary substantially by age group, academic position, and discipline, eliminating obstacles and promoting gender equality requires different strategies for different segments of academic science. What works in male-dominated disciplines or for young scientists may not work in female-dominated disciplines or for older scientists. Second, the idea that the value of international collaboration is intrinsically linked to reward structures, academic promotions, and access to research funding needs to be much more clearly formulated in policy documents and public debates. Female scientists need to be more aware of "internationalization accumulative advantage," especially that this advantage derives from joint publishing rather than international mobility, which is generally more difficult for females. As more international collaboration tends to imply higher publishing rates (and higher citation rates), the awareness that research internationalization plays a stratifying role in science needs to be shared by all. Moreover, understanding of "what counts" in assessments, promotions, and research funding agencies needs to be more widespread. Finally, acknowledging the fundamental role of international research collaboration in science today, programs fostering this type of collaboration for female scientists via numerous successful policy schemes should become a national and institutional policy priority (such as, for instance, the ADVANCE program from the National Science Foundation).

## Acknowledgments

The authors gratefully acknowledge the support of the Ministry of Science and Higher Education through its Dialog grant 0022/DLG/2019/10 (RESEARCH UNIVERSITIES).

## Notes

1. There were 38,750 records referring to 32,937 unique authors (more than one occurrence in Database II was found for 4,452 people or 13.51% of unique authors). There were 9,931 records that referred to more than one person, where 3,679 (82.63%) occurred twice, 609 (13.68%) occurred three times, and 169 (3.68%) occurred four or more times. Therefore, for duplicated records, a clerical review was performed (as suggested in Herzog *et al*., 2007). Manual verification of all duplicate records revealed that 1,207 records (12.15% in terms of duplicated records and 3.11% of all integrated records) were incorrectly assigned to the same person. These records were deleted from the integrated database, yielding N = 37,081 records.
2. The ASJC discipline codes were described in the paper in the following manner: AGRI Agricultural and Biological Sciences; HUM Arts and Humanities; BIO Biochemistry, Genetics, and Molecular Biology; BUS Business, Management, and Accounting; CHEMENG Chemical Engineering; CHEM Chemistry; COMP Computer Science; DEC Decision Science; DENT Dentistry, EARTH Earth and Planetary Sciences; ECON Economics, Econometrics, and Finance; ENER Energy; ENG Engineering; ENVIR Environmental Science; IMMU Immunology and Microbiology; MATER Materials Science; MATH Mathematics; MED Medicine; NEURO Neuroscience; NURS Nursing; PHARM Pharmacology, Toxicology, and Pharmaceutics; PHYS Physics and Astronomy; PSYCH Psychology; SOC Social Sciences; VET Veterinary; DENT Dentistry; and HEALTH Health Professions. Non-STEM disciplines in our analysis include BUS, DENT, ECON, HEALTH, HUM, MED, PSYCH, SOC, and VET.

# Appendix

**Table A1.** Structure of the Sample, All Polish Internationally Visible University Professors by Gender, Age Group, and Academic Position, Presented with Column and Row Percentages.

| | Assistant professor | | | | | | Associate professor | | | | | | Full professor | | | | | | Total | | | | | |
|---|---|---|---|---|---|---|---|---|---|---|---|---|---|---|---|---|---|---|---|---|---|---|---|---|
| | Female | | | Male | | | Female | | | Male | | | Female | | | Male | | | Female | | | Male | | |
| | *N* | % col | % row | *N* | % col | % row | *N* | % col | % row | *N* | % col | % row | *N* | % col | % row | *N* | % col | % row | *N* | % col | % row | *N* | % col | % row |
| 26–30 | 246 | 3.6 | 46.8 | 280 | 3.8 | 53.2 | 0 | 0.0 | 0.0 | 0 | 0.0 | 0.0 | 0 | 0.0 | 0.0 | 0 | 0.0 | 0.0 | 246 | 2.3 | 46.8 | 280 | 1.9 | 53.2 |
| 31–35 | 1636 | 23.8 | 50.2 | 1621 | 21.8 | 49.8 | 23 | 0.8 | 32.4 | 48 | 1.0 | 67.6 | 0 | 0.0 | 0.0 | 1 | 0.0 | 100.0 | 1659 | 15.6 | 49.8 | 1670 | 11.2 | 50.2 |
| 36–40 | 1937 | 28.2 | 49.8 | 1956 | 26.3 | 50.2 | 219 | 7.7 | 32.6 | 453 | 9.8 | 67.4 | 2 | 0.2 | 22.2 | 7 | 0.2 | 77.8 | 2158 | 20.3 | 47.2 | 2416 | 16.2 | 52.8 |
| 41–45 | 1589 | 23.1 | 49.0 | 1652 | 22.2 | 51.0 | 664 | 23.5 | 40.4 | 981 | 21.3 | 59.6 | 25 | 2.8 | 24.8 | 76 | 2.6 | 75.2 | 2278 | 21.5 | 45.7 | 2709 | 18.1 | 54.3 |
| 46–50 | 820 | 11.9 | 46.4 | 946 | 12.7 | 53.6 | 695 | 24.6 | 43.2 | 914 | 19.8 | 56.8 | 57 | 6.3 | 26.6 | 157 | 5.5 | 73.4 | 1572 | 14.8 | 43.8 | 2017 | 13.5 | 56.2 |
| 51–55 | 361 | 5.2 | 45.9 | 425 | 5.7 | 54.1 | 523 | 18.5 | 40.9 | 757 | 16.4 | 59.1 | 112 | 12.3 | 27.7 | 292 | 10.1 | 72.3 | 996 | 9.4 | 40.3 | 1474 | 9.9 | 59.7 |
| 56–60 | 195 | 2.8 | 46.7 | 223 | 3.0 | 53.3 | 336 | 11.9 | 39.6 | 513 | 11.1 | 60.4 | 144 | 15.9 | 27.8 | 374 | 13.0 | 72.2 | 675 | 6.4 | 37.8 | 1110 | 7.4 | 62.2 |
| 61–65 | 80 | 1.2 | 25.0 | 240 | 3.2 | 75.0 | 257 | 9.1 | 31.5 | 558 | 12.1 | 68.5 | 227 | 25.0 | 23.2 | 753 | 26.2 | 76.8 | 564 | 5.3 | 26.7 | 1551 | 10.4 | 73.3 |
| 66–70 | 13 | 0.2 | 12.9 | 88 | 1.2 | 87.1 | 105 | 3.7 | 23.7 | 338 | 7.3 | 76.3 | 302 | 33.3 | 23.7 | 974 | 33.9 | 76.3 | 420 | 4.0 | 23.1 | 1400 | 9.4 | 76.9 |
| 71+ | 0 | 0.0 | 0.0 | 13 | 0.2 | 100.0 | 8 | 0.3 | 15.1 | 45 | 1.0 | 84.9 | 38 | 4.2 | 13.5 | 243 | 8.4 | 86.5 | 46 | 0.4 | 13.3 | 301 | 2.0 | 86.7 |
| **Total** | **6877** | **100.0** | **48.0** | **7444** | **100.0** | **52.0** | **2830** | **100.0** | **38.1** | **4607** | **100.0** | **61.9** | **907** | **100.0** | **24.0** | **2877** | **100.0** | **76.0** | **10,614** | **100.0** | **41.6** | **14,928** | **100.0** | **58.4** |

**Table A2.** Percentage Differences in International Collaboration; All Polish Internationally Visible University Professors; by Collaboration Intensity (Low, Medium, and High), Gender, and Discipline (in %).

| | | All scientists | | Scientists: | | | Scientists: | | | Scientists: | | |
|---|---|---|---|---|---|---|---|---|---|---|---|---|
| | | | | **minimum 1 collaborative article in individual publication portfolio (Low intensity collaboration)** | | | **> 50% collaborative articles in individual publication portfolio (Middle intensity collaboration)** | | | **> 75% collaborative articles in individual publication portfolio (High intensity collaboration)** | | |
| | | ***N*** | **% col** | ***N*** | **% col** | **% row** | ***N*** | **% col** | **% row** | ***N*** | **% col** | **% row** |
| AGRI | Female | 1444 | 53.4 | 710 | 52.1 | 49.2 | 88 | 46.6 | 6.1 | 44 | 59.5 | 3 |
| | Male | 1258 | 46.6 | 652 | 47.9 | 51.8 | 101 | 53.4 | 8 | 30 | 40.5 | 2.4 |
| BIO | Female | 1068 | 60 | 645 | 56 | 60.4 | 119 | 55.1 | 11.1 | 48 | 53.3 | 4.5 |
| | Male | 712 | 40 | 507 | 44 | 71.2 | 97 | 44.9 | 13.6 | 42 | 46.7 | 5.9 |
| CHEM | Female | 756 | 51.3 | 514 | 48.8 | 68 | 103 | 48.8 | 13.6 | 34 | 43.6 | 4.5 |
| | Male | 719 | 48.7 | 539 | 51.2 | 75 | 108 | 51.2 | 15 | 44 | 56.4 | 6.1 |
| CHEMENG | Female | 185 | 38.5 | 73 | 44.2 | 39.5 | 7 | 38.9 | 3.8 | 5 | 38.5 | 2.7 |
| | Male | 296 | 61.5 | 92 | 55.8 | 31.1 | 11 | 61.1 | 3.7 | 8 | 61.5 | 2.7 |
| COMP | Female | 170 | 16.5 | 61 | 15.6 | 35.9 | 18 | 13.8 | 10.6 | 13 | 17.3 | 7.6 |
| | Male | 860 | 83.5 | 331 | 84.4 | 38.5 | 112 | 86.2 | 13 | 62 | 82.7 | 7.2 |
| DEC | Female | 24 | 44.4 | 6 | 46.2 | 25 | 4 | 80 | 16.7 | 4 | 80 | 16.7 |
| | Male | 30 | 55.6 | 7 | 53.8 | 23.3 | 1 | 20 | 3.3 | 1 | 20 | 3.3 |
| EARTH | Female | 385 | 33.4 | 149 | 30.9 | 38.7 | 40 | 31.3 | 10.4 | 28 | 43.1 | 7.3 |
| | Male | 769 | 66.6 | 333 | 69.1 | 43.3 | 88 | 68.8 | 11.4 | 37 | 56.9 | 4.8 |

(*Continued*)

**Table A2.** *Continued.*

| | | All scientists | | Scientists: | | | Scientists: | | | Scientists: | | |
|---|---|---|---|---|---|---|---|---|---|---|---|---|
| | | | | **minimum 1 collaborative article in individual publication portfolio (Low intensity collaboration)** | | | **> 50% collaborative articles in individual publication portfolio (Middle intensity collaboration)** | | | **> 75% collaborative articles in individual publication portfolio (High intensity collaboration)** | | |
| | | ***N*** | **% col** | ***N*** | **% col** | **% row** | ***N*** | **% col** | **% row** | ***N*** | **% col** | **% row** |
| ENER | Female | 82 | 27.8 | 17 | 19.8 | 20.7 | 5 | 20.8 | 6.1 | 4 | 25 | 4.9 |
| | Male | 213 | 72.2 | 69 | 80.2 | 32.4 | 19 | 79.2 | 8.9 | 12 | 75 | 5.6 |
| ENG | Female | 501 | 14.9 | 133 | 13 | 26.5 | 26 | 14 | 5.2 | 10 | 10.9 | 2 |
| | Male | 2857 | 85.1 | 894 | 87 | 31.3 | 160 | 86 | 5.6 | 82 | 89.1 | 2.9 |
| ENVIR | Female | 848 | 50.5 | 347 | 49.2 | 40.9 | 36 | 38.3 | 4.2 | 16 | 38.1 | 1.9 |
| | Male | 832 | 49.5 | 358 | 50.8 | 43 | 58 | 61.7 | 7 | 26 | 61.9 | 3.1 |
| IMMU | Female | 90 | 75.6 | 57 | 74 | 63.3 | 14 | 87.5 | 15.6 | 9 | 90 | 10 |
| | Male | 29 | 24.4 | 20 | 26 | 69 | 2 | 12.5 | 6.9 | 1 | 10 | 3.4 |
| MATER | Female | 495 | 33.9 | 309 | 33.7 | 62.4 | 67 | 33.7 | 13.5 | 29 | 35.8 | 5.9 |
| | Male | 967 | 66.1 | 608 | 66.3 | 62.9 | 132 | 66.3 | 13.7 | 52 | 64.2 | 5.4 |
| MATH | Female | 259 | 25.2 | 130 | 23.2 | 50.2 | 44 | 23.2 | 17 | 13 | 17.1 | 5 |
| | Male | 767 | 74.8 | 430 | 76.8 | 56.1 | 146 | 76.8 | 19 | 63 | 82.9 | 8.2 |
| PHARM | Female | 169 | 66.5 | 96 | 61.5 | 56.8 | 14 | 60.9 | 8.3 | 3 | 50 | 1.8 |
| | Male | 85 | 33.5 | 60 | 38.5 | 70.6 | 9 | 39.1 | 10.6 | 3 | 50 | 3.5 |
| PHYS | Female | 182 | 16.6 | 133 | 16.3 | 73.1 | 46 | 13.9 | 25.3 | 25 | 14.5 | 13.7 |
| | Male | 916 | 83.4 | 685 | 83.7 | 74.8 | 284 | 86.1 | 31 | 147 | 85.5 | 16 |
| BUS | Female | 372 | 52.1 | 125 | 55.6 | 33.6 | 56 | 54.4 | 15.1 | 43 | 53.8 | 11.6 |
| | Male | 342 | 47.9 | 100 | 44.4 | 29.2 | 47 | 45.6 | 13.7 | 37 | 46.3 | 10.8 |

(*Continued*)

**Table A2.** *Continued.*

| | | All scientists | | Scientists: | | | Scientists: | | | Scientists: | | |
|---|---|---|---|---|---|---|---|---|---|---|---|---|
| | | | | **minimum 1 collaborative article in individual publication portfolio (Low intensity collaboration)** | | | **> 50% collaborative articles in individual publication portfolio (Middle intensity collaboration)** | | | **> 75% collaborative articles in individual publication portfolio (High intensity collaboration)** | | |
| | | ***N*** | **% col** | ***N*** | **% col** | **% row** | ***N*** | **% col** | **% row** | ***N*** | **% col** | **% row** |
| DENT | Female | 57 | 76 | 11 | 78.6 | 19.3 | 0 | 0 | 0 | 0 | 0 | 0 |
| | Male | 18 | 24 | 3 | 21.4 | 16.7 | 0 | 0 | 0 | 0 | 0 | 0 |
| ECON | Female | 186 | 49.1 | 36 | 44.4 | 19.4 | 13 | 56.5 | 7 | 10 | 62.5 | 5.4 |
| | Male | 193 | 50.9 | 45 | 55.6 | 23.3 | 10 | 43.5 | 5.2 | 6 | 37.5 | 3.1 |
| HEALTH | Female | 23 | 34.3 | 3 | 14.3 | 13 | 0 | 0 | 0 | 0 | 0 | 0 |
| | Male | 44 | 65.7 | 18 | 85.7 | 40.9 | 1 | 100 | 2.3 | 0 | 0 | 0 |
| HUM | Female | 527 | 49.8 | 44 | 41.1 | 8.3 | 18 | 41.9 | 3.4 | 17 | 43.6 | 3.2 |
| | Male | 531 | 50.2 | 63 | 58.9 | 11.9 | 25 | 58.1 | 4.7 | 22 | 56.4 | 4.1 |
| MED | Female | 1920 | 53.7 | 934 | 49.8 | 48.6 | 110 | 48.9 | 5.7 | 44 | 46.3 | 2.3 |
| | Male | 1654 | 46.3 | 941 | 50.2 | 56.9 | 115 | 51.1 | 7 | 51 | 53.7 | 3.1 |
| PSYCH | Female | 194 | 63.8 | 72 | 54.5 | 37.1 | 28 | 53.8 | 14.4 | 16 | 57.1 | 8.2 |
| | Male | 110 | 36.2 | 60 | 45.5 | 54.5 | 24 | 46.2 | 21.8 | 12 | 42.9 | 10.9 |
| SOC | Female | 494 | 49.8 | 105 | 47.5 | 21.3 | 31 | 38.8 | 6.3 | 23 | 41.1 | 4.7 |
| | Male | 498 | 50.2 | 116 | 52.5 | 23.3 | 49 | 61.3 | 9.8 | 33 | 58.9 | 6.6 |
| VET | Female | 146 | 44 | 87 | 40.8 | 59.6 | 1 | 25 | 0.7 | 0 | 0 | 0 |
| | Male | 186 | 56 | 126 | 59.2 | 67.7 | 3 | 75 | 1.6 | 1 | 100 | 0.5 |
| Total | Female | 10,577 | 41.5 | 4797 | 40.5 | 45.4 | 888 | 35.7 | 8.4 | 438 | 36.2 | 4.1 |
| | Male | 14,886 | 58.5 | 7057 | 59.5 | 47.4 | 1602 | 64.3 | 10.8 | 772 | 63.8 | 5.2 |

## Supporting Information

Additional supporting information may be found online in the Supporting Information section at the end of the article.

**Table A1**: Structure of the sample, all Polish internationally productive university professors by gender, age group, and academic position, presented with column and row percentages.
**Table A2**: Percentage differences in international collaboration; all Polish internationally productive university professors; by collaboration intensity (low, medium, and high), gender, and discipline (in %).